\documentclass{aa}
\usepackage{txfonts}
\usepackage{graphicx}
\usepackage{lscape}
\usepackage{natbib}
\bibpunct{(}{)}{;}{a}{}{,} 
\newcommand{\little}{\fontsize{8}{9}\selectfont} 
\begin{document}
\title{Molecular abundances in carbon-rich circumstellar envelopes}
\author{P. M. Woods\inst{1,2}, F. L. Sch\"oier\inst{3}, L.-\AA. Nyman\inst{1,4}, H. Olofsson\inst{5}}
\titlerunning{Molecular abundances in carbon-rich CSEs}
\authorrunning{Woods et al.}
\offprints{PMW, \email{pwoods@eso.org}}
\institute{European Southern Observatory, Alonso de Cordova 3107, Casilla 19001, Santiago 19, Chile.
\and
UMIST, Department of Physics, P.O. Box 88, Manchester M60 1QD, UK.
\and
Leiden Observatory, P.O. Box 9513, 2300 RA Leiden, The Netherlands.
\and
Onsala Space Observatory, 439 92 Onsala, Sweden.
\and
Stockholm Observatory, 133 36 Saltsj\"obaden, Sweden.
}
%
\date{A\&A accepted}
\abstract{A millimetre molecular line survey of seven high
mass-loss rate carbon stars in both the northern and southern skies is
presented.  A total of 196 emission lines (47 transitions) from 24
molecular species were detected. The observed CO emission is used to
determine mass-loss rates and the physical structure of the
circumstellar envelope, such as the density and temperature structure,
using a detailed radiative transfer analysis. This enables abundances
for the remaining molecular species to be determined. The derived
abundances generally vary between the sources by no more than a factor
of five indicating that circumstellar envelopes around carbon stars
with high mass-loss rates have similar chemical compositions. However,
there are some notable exceptions.  The most striking difference
between the abundances are reflecting the spread in the
$^{12}$C/$^{13}$C-ratio of about an order of magnitude between the
sample stars, which mainly shows the results of nucleosynthesis.
The abundance of SiO also shows a variation of more than an order of
magnitude between the sources and is on the average more than an order
of magnitude more abundant than predicted from photospheric chemistry
in thermal equilibrium. The over-abundance of SiO is consistent with
dynamical modelling of the stellar atmosphere and the inner parts of
the wind where a pulsation-driven shock has passed. This scenario is
possibly further substantiated by the relatively low amount of CS
present in the envelopes. The chemistry occurring in the outer
envelope is consistent with current photochemical models.
}
\maketitle

\section{Introduction}

The chemistry associated with carbon stars has long been known to be
rich and complex in comparison to the alternative O-rich regime (i.e.,
where C/O\,$<$\,1).  This is in part due to the favourable bonding of
the carbon atom, enabling long chains and complex species to form.
Most of the current understanding of carbon stars has come from both
observational and theoretical work on the high-mass-losing carbon
star, \object{IRC+10216}.  This source, which lies within 200\,pc and
presents an ideal specimen for the study of carbon-rich envelopes, has
been mapped interferometrically in various molecular species
\citep[e.g.][]{BiegingTafalla1993, DayalBieging1993, DayalBieging1995,
Gensheimer_etal1995, Guelin_etal1993, Guelin_etal1996, Lucas_etal1995,
LucasGuelin1999} and has had models of its dust \citep[a good summary
is given by][]{Menshchikov_etal2001} and chemistry
\citep[e.g.][]{Millar_etal2000} constructed.  These tools have
produced groundbreaking results and have been used to set a paradigm
for what has come to be known as \textquotedblleft carbon
chemistry\textquotedblright\ in connection with evolved stars.
However, the accuracy of employing \object{IRC+10216} chemistry to
similar carbon stars has been little-tested due to the difficulties in
observing them.  Much work has been done on the carbon-rich post-AGB
sources \object{CRL 618} and \object{CRL 2688}, and the chemistry of
\object{CRL 618} in particular has been modelled by
\citet{Woods_etal2002}.  Detailed chemical studies of carbon stars on
the AGB have been few in number, but examples include the molecular
line survey of \object{IRAS 15194--5115}, a peculiar $^{13}$C-rich
star \citep{Nyman_etal1993}.  Carbon star surveys which include
molecular-line comparisons are fewer, and have been limited in the
number of lines observed. \citet{Olofsson_etal1993b} detected some 40
stars in a handful of species other than CO. The sample of
\citet{Bujarrabal_etal1994} included 16 carbon stars, with up to ten
molecular lines observed in each.  A more recent survey by
\citet{Olofsson_etal1998} detected 22 carbon stars in up to 6
molecular lines.

\begin{table*}
   \caption{Positions, luminosities, periods and calculated distances of the sample of carbon stars.}
\label{stellardata}
   \begin{flushleft}
   \begin{tabular}{lllcccccc}
   \hline\hline
 IRAS No.    & Other cat. name & B1950 Co-ords.   & $P$    & $L$      & $D$     & $T_*$   & $T_{\mathrm{d}}$ &
$L_{\mathrm{d}}$/$L_*$\\
             &                 &                  & [days] & [L$_{\odot}$]  & [pc]    & [K]     & [K]              \\
    \hline
$07454-7112$ & \object{AFGL\,4078}     & 07:45:25.7 $-$71:12:18   &  ---       &  9\,000$^a$                   & 710  & 1\,200
&  \phantom{0}710 & 4.3\\
$09452+1330$ & \object{IRC+10216}      & 09:45:15.0 $+$13:30:45  & 630     & 9\,600\phantom{$^b$} & 120 & ---      &
\phantom{0}510 & ---\\
$10131+3049$ & \object{CIT\,6}         & 10:13:11.5 $+$30:49:17  & 640     & 9\,700\phantom{$^b$} & 440 & 1\,300 &
\phantom{0}510 & 6.7\\
$15082-4808$ & \object{AFGL\,4211}     & 15:08:13.0 $-$48:08:43   & ---        & 9\,000$^a$                     & 640 & ---
&  \phantom{0}590 & ---\\
$15194-5115$ & ---                     & 15:19:26.9 $-$51:15:19   & 580     & 8\,800\phantom{$^b$} & 600 &  \phantom{0}930 &
\phantom{0}480 & 2.2\\
$23166+1655$ & \object{AFGL\,3068}     & 23.16.42.4 $+$16:55:10  & 700     & 7\,800\phantom{$^b$} & 820 & --- & 1\,000 & --- \\
$23320+4316$ & \object{IRC+40540}      & 23:32:00.4 $+$43:16:17  & 620     & 9\,400\phantom{$^b$} & 630 & 1\,100 &
\phantom{0}610 & 6.6\\
  \hline
   \end{tabular}
%
  {\footnotesize
   \begin{enumerate}
        \renewcommand{\labelenumi}{(\alph{enumi})}
        \item Assumed value.
   \end{enumerate}
     }
\end{flushleft}
\end{table*}

The survey work presented here purports to be the most complete and
consistent molecular-line survey in AGB carbon stars to date, covering
high mass-loss rate objects in both the northern and the
southern sky.  Previously unpublished spectra of five stars
(\object{IRAS 15082--4808}, \object{IRAS 07454--7112}, \object{CIT 6},
\object{AFGL 3068} and \object{IRC+40540}) are presented, and spectra
taken towards \object{IRC+10216} and \object{IRAS 15194--5115} with
the Swedish-ESO Submillimetre Telescope
\citep[SEST;][]{Nyman_etal1993} and \object{IRC+10216} with the Onsala
Space Observatory (OSO) 20\,m telescope are used to supplement the
survey.  Comparison of data from \object{IRC+10216} taken with both
telescopes affords a high degree of confidence in the relative
calibration that can be derived.

Up to 51 molecular lines were observed in the sample of 7
high-mass-losing carbon stars, of which 47 were clearly
detected. Mass-loss rates, dust properties and the
$^{12}$CO/$^{13}$CO-ratio are calculated self-consistently using a
radiative transfer method \citep{SchoierOlofsson2000,
SchoierOlofsson2001, Schoier_etal2002}.  An approach similar to that
of \citet{Nyman_etal1993} is adopted to calculate fractional
abundances (including upper limits), and a detailed analysis of the
comparison between the calculated abundances is carried out.  Hence,
Sect.~\ref{obs} details the observations carried out and the
instrumentation used.  Section~\ref{results} gives the observational
results, including a presentation of various spectra.
Section~\ref{modelling} details the NLTE radiative transfer code used
to determine the envelope parameters and Sect.~\ref{abundances}
explains the method of calculating chemical abundances.  The results
and deductions are discussed in Sect.~\ref{discuss}.

\section{Observations}
\label{obs}
\subsection{Carbon star sample}

Following the CO survey of \citet{Nyman_etal1992}, several carbon
stars, which were bright in CO lines, were selected for a more
comprehensive molecular line search. These stars are rare in that
they are all losing mass at a very high rate, and hence are more
likely to produce strong emission from a variety of molecular lines.
The sample of seven carbon stars (Table~\ref{stellardata}) was
observed using both the 15\,m Swedish--ESO Submillimetre Telescope
\citep{Booth_etal1989} during the period 1987 -- 1996, and the Onsala
20\,m telescope, in 1994. The SEST, situated on La Silla, Chile, was
used to observe \object{IRAS 07454-7112}, \object{IRAS 15082-4808},
and \object{IRAS 15194-5115}. The 20\,m telescope, located at the
Onsala Space Observatory (OSO) in Sweden, observed the remaining three
sources, \object{CIT 6}, \object{AFGL 3068} and \object{IRC
+40540}. Both telescopes were used to observe the well-studied carbon
star, \object{IRC +10216} in order to determine the relative
calibration between the two telescopes.

The JCMT public archive\footnote{{\tt
http://www.jch.hawaii.edu/JACpublic/JCMT/} \\ The JCMT is operated by
the Joint Astronomy Centre in Hilo, Hawaii on behalf of the present
organisations: the Particle Physics and Astronomy Research Council in
the United Kingdom, the National Research Council of Canada and the
Netherlands Organization for Scientific Research.}  was searched for
complementary line observations, in particular the CO lines used in
the radiation transfer modelling described in Sect.~\ref{modelling}.
Lines for which multi-epoch observations are available in the JCMT
archive typically display intensities that are consistent to
$\sim$20\% \citep{SchoierOlofsson2001}. In addition, interferometric
observations of the CO($J$=1--0) brightness distribution around some
of the sample stars have been performed \citep{Neri98} using the
Plateu de Bure interferometer (PdBI), France.  The data are publically
available and have been used in this paper.

The coordinates used for each individual source are listed in
Table~\ref{stellardata}.  Also shown in Table~\ref{stellardata} are
the adopted luminosities and distances to be used in the molecular
excitation analysis.  For stars where a period has been determined
(see Table~\ref{stellardata}) the period-luminosity relation from
\citet{GroenWhitelock1996} was used to estimate the corresponding
luminosity.  If a reliable period is not available the total
bolometric luminosity was fixed to 9\,000\,L$_{\sun}$.  The distance
was then obtained from the luminosity using the observed bolometric
magnitude.  \citet{SchoierOlofsson2001} used the same approach when
determining the distances to a large sample of optically bright carbon
stars and concluded that there were no apparent systematic effect when
comparing with estimates based upon Hipparcos parallaxes, although the
scatter is large and the distance estimate for an individual source is
subject to some uncertainty of up to a factor of $\sim$2. The effects
of the adopted distance on the molecular excitation will be addressed
in Sect.~\ref{distance}.

If, for simplicity, the central ratiation field is represented by one
or two blackbodies their properties can be determined from a fit to
the observed spectral energy distribution (SED) as described in
\citet{Kerschbaum1999}. A fit to the SED gives the two blackbody
temperatures ($T_*$ and $T_{\mathrm{d}}$) as well as their relative
luminosities ($L_{\mathrm{d}}$/$L_*$).  The parameters obtained in this
fashion are presented in Table~\ref{stellardata}.

\subsection{Instrumentation}

\begin{table}
\begin{flushleft}
\caption{Beam widths and efficiencies at selected frequencies.}
\label{efficiencies}
\begin{tabular}{c cc c cc}
\hline\hline
          &\multicolumn{2}{c}{SEST 15\,m}&&\multicolumn{2}{c}{Onsala 20\,m}\\
\cline{2-3}\cline{5-6}
Frequency &\multicolumn{1}{c}{FWHM}& $\eta_{\rm mb}$ &&\multicolumn{1}{c}{FWHM}& $\eta_{\rm mb}$\\
{\rm [GHz]} &\multicolumn{1}{c}{[$\arcsec$]}& &&\multicolumn{1}{c}{[$\arcsec$]}& \\
\hline
\phantom{0}86&	57	&	0.75	&&	44	&	0.58	\\
100	&	51	&	0.73	&&	39	&	0.53	\\
115	&	45	&	0.70	&&	33	&	0.48	\\
230	&	23	&	0.50	& & --- & --- \\				
265	&	21	&	0.42	& & --- & --- \\				
\hline
\end{tabular}
\end{flushleft}
\end{table}

The SEST is equipped with two acousto-optical spectrometers (HRS,
86\,MHz bandwidth with 43\,kHz channel separation and 80\,kHz
resolution; LRS, 500\,MHz bandwidth with 0.7\,MHz channel separation
and a resolution of 1.4\,MHz). The receivers used were dual
polarization Schottky receivers at both 3 and 1.3\,mm
wavelength. Typical system temperatures above the atmosphere were
400--500\,K and 1000-1800\,K, respectively.

The OSO 20\,m telescope uses two filterbanks (MUL B, 64\,MHz bandwidth with a
channel width of 250\,kHz; MUL A, 512\,MHz bandwidth and a channel width
of 1\,MHz). The receiver used was a horizontally, linearly polarised
SIS receiver with a typical system temperature of 400--500\,K above
the atmosphere.

All observations were performed using the dual beam switching method,
which places the source alternately in two beams, and yields very flat
baselines. Beam separation was in both cases about 11\farcm5.
Calibration was done with the standard chopper-wheel method.  The
intensity scales of the spectra are given in main-beam brightness
temperature (the corrected antenna temperature, $T_{\mathrm{A}}^*$,
divided by the main-beam efficiency, $\eta_{\mathrm{mb}}$).  Main-beam
efficiencies and FWHM beam widths are given in Table
\ref{efficiencies}\ for both telescopes.

\section{Observational results}
\label{results}

   \begin{figure*}
   \centering{   
   \includegraphics[width=16.75cm]{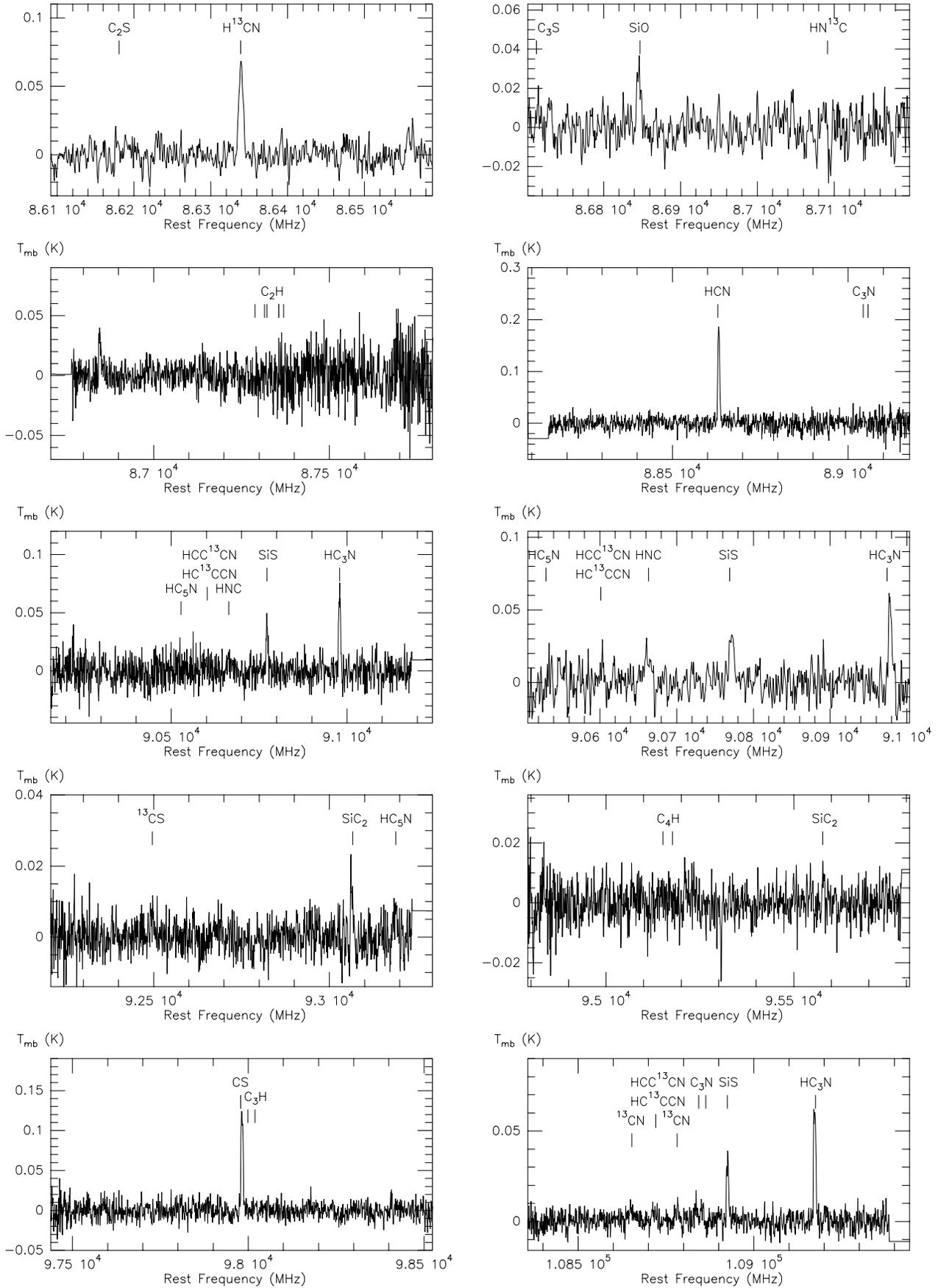}
   \caption{Low-resolution spectra of \object{IRAS07454--7112}, obtained
   with the SEST.}
   \label{07454lrs1}}
   \end{figure*}

   \begin{figure*}
   \centering{   
   \includegraphics[width=16.75cm]{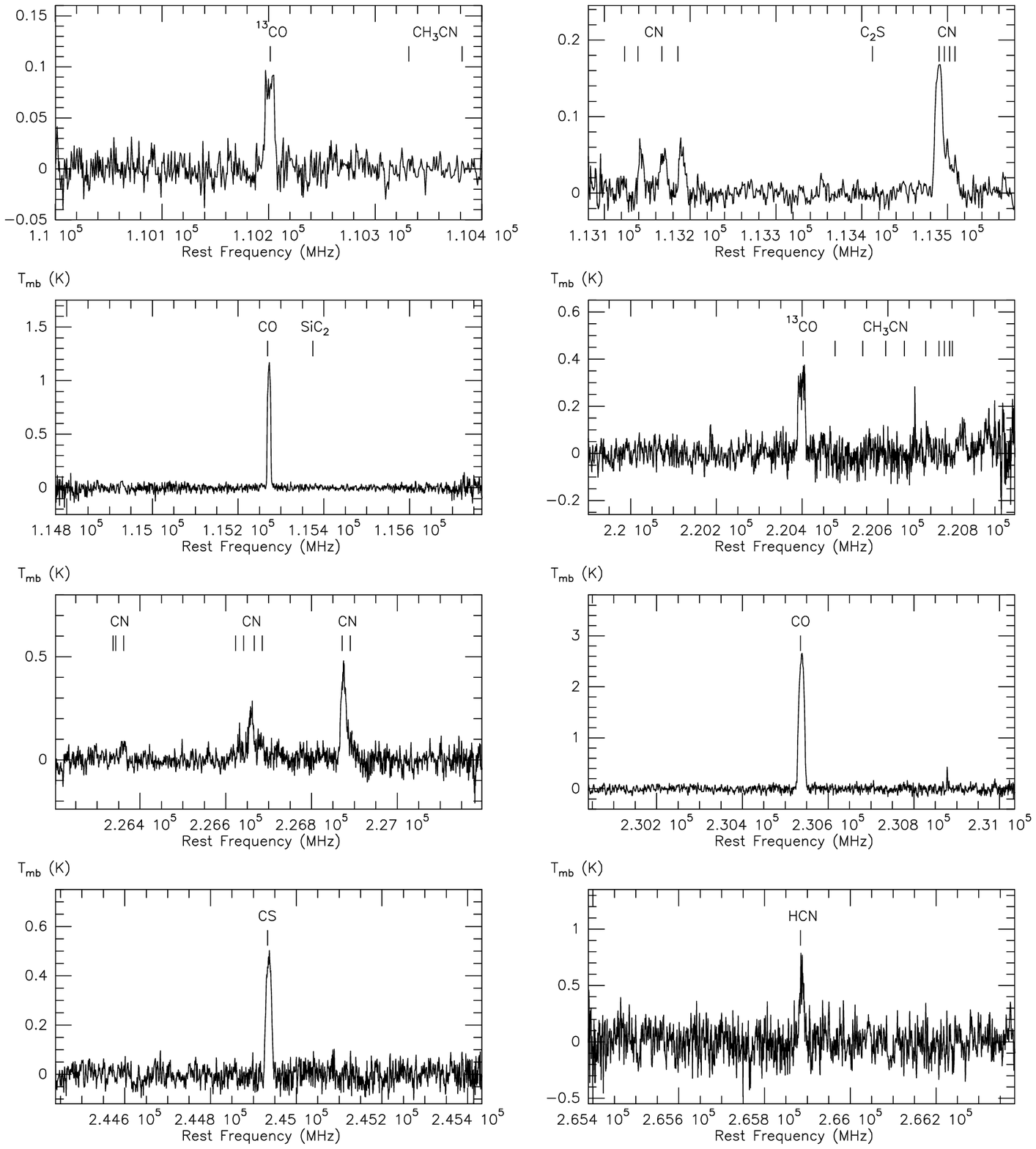}
   \caption{Low-resolution spectra of \object{IRAS07454--7112}, obtained
   with the SEST.}
   \label{07454lrs2}}
   \end{figure*}

\subsection{Observed lines}

A total of 196 lines were detected in the sample.  47 transitions of
24 molecular species were detected, and upper limits for another 95
relevant transitions were also obtained, including another three
species (C$_2$S, C$_3$S and SO).  Previously unpublished spectra are
shown in Figs.~\ref{07454lrs1}--\ref{07454lrs2} \&
\ref{07454hrs1}--\ref{40540lrs1}.  Table \ref{tabdetlines}\ lists the
detections in all seven sources, together with their peak and
integrated intensities.  If a line has a hyperfine structure, the
frequency and intensity of the strongest component is listed, and the
integrated intensity is the sum over all hyperfine components.  Values
of $T_{\mathrm{mb}}$\ for lines where no detection was made are the
rms noise values.

Almost all lines observed in \object{IRC+10216} by the SEST were
observed in the same source with the OSO 20\,m telescope.  The
majority of those which were not are due to the lack of a 1.3\,mm
receiver at OSO. A large proportion of these lines were observed in
the remaining six sources.  The purpose of observing
\object{IRC+10216} twice, with different telescopes, was to ascertain
the relative calibration between the two different setups, and hence
gain a basis from which good comparisons could be made between the
entire sample of carbon stars.

The $^{13}$CS ($J$=2--1) line observed in some of these sources is
partially blended with the C$_3$S ($J$=16--15) line
\citep{Kahane_etal1988}. The $^{13}$CS ($J$=2--1) integrated
intensities include both these lines. It is assumed that the relative
intensities of these two lines are constant in all the objects, and
hence will not affect comparisons. The lines of two HC$_3$N isotopes,
HC$^{13}$CCN and HCC$^{13}$CN, lie very closely and are in all cases
blended together. No attempt is made to separate the components.

\subsection{Line profiles}

There are four characteristic line profile shapes which give
information on the source being observed. These lineshapes are most
typically seen in CO emission since it is often strongest \citep[for a
comprehensive review of line profiles see][]{Olofsson_etal1993}. For
optically thick emission the line profile can be described as
parabolic for unresolved sources or flat-topped for resolved
sources. A parabolic profile is shown by most of the $^{12}$CO
emission lines in the sample except in \object{IRC+10216} and
\object{IRAS 15194--5115}, where the CO ($J$=1--0) emission shows a
flat-topped profile.

When emission is optically thin, a flat-topped profile is seen for
unresolved sources, and a double-peaked profile for a resolved
source. Examples of these profiles can be seen in the $^{13}$CO
emission towards \object{IRAS 15082--4808} and \object{IRC+10216}.
In the following calculations, all lines are assumed optically thin,
except those from CO and HCN.

The values of the expansion velocity, $v_{\mathrm{exp}}$, of the
circumstellar envelopes quoted in Table~\ref{tabmlr} are obtained from
the radiative transfer CO modelling of these sources (presented in
Sect.~\ref{modelling}) where its value is adjusted until a good fit
to the observed line profiles is obtained.  No trends in the widths of
the observed lines are present which lends further support to the
assumption adopted here that these envelopes are expanding at constant
velocities.

\begin{landscape}
\begin{table}
\begin{flushleft}
\caption{Detected lines}
\label{tabdetlines}
\little
\begin{tabular}{l l r r rr r rr r rr r rr}
\hline\hline
Molecule & Transition & Frequency &&
\multicolumn{2}{c}{IRC+10216 (SEST)}&&\multicolumn{2}{c}{IRC+10216 (OSO)}&&\multicolumn{2}{c}{IRAS15194-5115}&&\multicolumn{2}{c}{IRAS15082-4808}\\
 \cline{5-6}\cline{8-9}\cline{11-12}\cline{14-15}
 & & && $T_{\rm mb}$ & $\int T_{\rm mb}{\rm d}v$ && $T_{\rm mb}$ & $\int T_{\rm mb}{\rm d}v$ && $T_{\rm mb}$ & $\int T_{\rm mb}{\rm d}v$ && $T_{\rm mb}$ & $\int T_{\rm mb}{\rm d}v$\\
 & & [MHz] && [K] & [K\,kms$^{-1}$] && [K] & [K\,kms$^{-1}$] && [K] & [K\,kms$^{-1}$] && [K] & [K\,kms$^{-1}$] \\ 
\hline
HC$_5$N      & J=32--31               & \phantom{0}85201.348           && 0.15       & 4.78      && ---       & ---       && $<$\,0.01  & $<$\,0.13 && $<$\,0.01  & $<$\,0.10 \\
C$_3$H$_2$   & 2$_{1,2}$--1$_{0,1}$   & \phantom{0}85338.905           && 0.11       & 2.49      && ---       & ---       && 0.02       & 0.84      && $<$\,0.01  & $<$\,0.10 \\
C$_4$H       & N=9--8, J=19/2--17/2   & \phantom{0}85634.00\phantom{0} && 0.20       & 9.60      && ---       & ---       && ---        & ---       && $<$\,0.01  & $<$\,0.05 \\
C$_4$H       & N=9--8, J=17/2--15/2   & \phantom{0}85672.57\phantom{0} && 0.20       & 9.60      && ---       & ---       && ---        & ---       && $<$\,0.01  & $<$\,0.05 \\
C$_2$S       & 6(7)--5(6)             & \phantom{0}86181.413           && $<$\,0.03  & $<$\,0.46 && $<$\,0.04 & $<$\,0.76 && $<$\,0.07  & $<$\,0.11 && $<$\,0.04  & $<$\,0.06 \\
H$^{13}$CN   & J=1--0                 & \phantom{0}86340.184           && 3.23       & 87.20     && 3.86      & 114.14    && 0.57       & 20.93     && 0.09       & 3.06      \\
C$_3$S       & J=15--14               & \phantom{0}86708.379           && $<$\,0.02  & $<$\,0.30 && $<$\,0.03 & $<$\,0.65 && $<$\,0.01  & $<$\,0.08 && $<$\,0.01  & $<$\,0.06 \\
SiO          & J=2--1            & \phantom{0}86846.998           && 0.71       & 20.40     && 0.95      & 24.48     && 0.16       & 5.33      && 0.07       & 1.84      \\
HN$^{13}$C   & J=1--0                 & \phantom{0}87090.859           && $<$\,0.04  & $<$\,0.62 && ---       & ---       && 0.02       & 1.11      && $<$\,0.01  & $<$\,0.11 \\
C$_2$H       & N=1--0                 & \phantom{0}87316.925           && 0.52       & 41.60     && 0.66      & 49.48     && 0.13       & 11.50     && 0.04       & 4.04      \\
HCN          & J=1--0                 & \phantom{0}88631.847           && 7.80       & 190.93    && 9.14      & 214.04    && 0.55       & 15.33     && 0.37       & 11.26     \\
C$_3$N       & N=9--8, J=19/2--17/2   & \phantom{0}89045.59\phantom{0} && 0.31       & 9.19      && ---       & ---       && $<$\,0.01  & $<$\,0.07 && $<$\,0.01  & $<$\,0.09 \\
C$_3$N       & N=9--8, J=17/2--15/2   & \phantom{0}89064.36\phantom{0} && 0.31       & 9.19      && ---       & ---       && $<$\,0.01  & $<$\,0.07 && $<$\,0.01  & $<$\,0.09 \\
HC$_5$N      & J=34--33               & \phantom{0}90525.892           && 0.16       & 2.58      && $<$\,0.03 & $<$\,0.67 && ---        & ---       && $<$\,0.01  & $<$\,0.10 \\
HC$^{13}$CCN & J=10--9                & \phantom{0}90593.059           && 0.09       & 1.98      && $<$\,0.03 & $<$\,0.34 && 0.02       & 0.76      && $<$\,0.01  & $<$\,0.05 \\
HCC$^{13}$CN & J=10--9                & \phantom{0}90601.791           && 0.09       & 1.98      && $<$\,0.03 & $<$\,0.34 && 0.02       & 0.76      && $<$\,0.01  & $<$\,0.05 \\
HNC          & J=1--0                 & \phantom{0}90663.543           && 0.74       & 26.49     && 0.58      & 17.22     && 0.08       & 3.72      && 0.05       & 1.70      \\
SiS          & J=5--4                 & \phantom{0}90771.546           && 1.14       & 35.54     && 1.07      & 39.47     && 0.05       & 2.78      && 0.03       & 1.58      \\
HC$_3$N      & J=10--9                & \phantom{0}90978.993           && 2.42       & 64.59     && 2.00      & 52.63     && 0.09       & 3.71      && 0.11       & 4.15      \\
$^{13}$CS    & J=2--1                 & \phantom{0}92494.299           && 0.07       & 2.97      && ---       & ---       && 0.06       & 2.67      && $<$\,0.01  & $<$\,0.10 \\
SiC$_2$      & 4$_{0,4}$--3$_{0,3}$   & \phantom{0}93063.639           && 0.46       & 14.60     && ---       & ---       && ---        & ---       && 0.03       & 0.40      \\
HC$_5$N      & J=35--34               & \phantom{0}93188.127           && 0.12       & 5.50      && ---       & ---       && ---        & ---       && $<$\,0.01  & $<$\,0.10 \\
C$_4$H       & N=10--9, J=21/2--19/2  & \phantom{0}95150.32\phantom{0} && 0.19       & 6.08      && 0.18      & 8.75      && 0.04       & 1.95      && 0.02       & 0.47      \\
C$_4$H       & N=10--9, J=19/2--17/2  & \phantom{0}95188.94\phantom{0} && 0.19       & 6.08      && 0.18      & 8.75      && 0.04       & 1.95      && 0.02       & 0.47      \\
SiC$_2$      & 4$_{2,2}$--3$_{2,1}$   & \phantom{0}95579.389           && 0.32       & 10.73     && ---       & ---       && ---        & ---       && $<$\,0.01  & $<$\,0.08 \\
C$^{34}$S    & J=2--1                 & \phantom{0}96412.982           && 0.15       & 5.66      && ---       & ---       && ---        & ---       && ---        & ---       \\
CS           & J=2--1                 & \phantom{0}97980.968           && 2.92       & 81.23     && 4.06      & 118.89    && 0.37       & 15.21     && 0.25       & 8.44      \\
C$_3$H       & $^2\Pi_{1/2}$,9/2--7/2 & \phantom{0}97995.450           && 0.18       & 11.27     && 0.28      & 14.54     && 0.01       & 0.64      && $<$\,0.01  & $<$\,0.11 \\
SO           & J=3--2                 & \phantom{0}99299.879           && $<$\,0.02  & $<$\,0.25 && ---       & ---       && ---        & ---       &&  ---       & ---      \\
HC$_3$N      & J=11--10               & 100076.389                     && 2.21       & 53.84     && ---       & ---       && ---        & ---       &&  ---       & ---      \\
HC$^{13}$CCN & J=12--11               & 108710.523                     && 0.09       & 2.34      && 0.14      & 1.22      && 0.01       & 0.39      && $<$\,0.01  & $<$\,0.03 \\
HCC$^{13}$CN & J=12--11               & 108721.008                     && 0.09       & 2.34      && 0.14      & 1.22      && 0.01       & 0.39      && $<$\,0.01  & $<$\,0.03 \\
$^{13}$CN    & N=1--0                 & 108780.201                     && 0.11       & 8.56      && 0.24      & 6.07      && 0.02       & 2.18      && $<$\,0.01  & $<$\,0.05 \\
C$_3$N       & N=11--10, J=23/2--21/2 & 108834.27\phantom{0}           && 0.39       & 11.76     && 0.64      & 18.00     && 0.01       & 0.68      && 0.02       & 0.63      \\
C$_3$N       & N=11--10, J=21/2--19/2 & 108853.02\phantom{0}           && 0.39       & 11.76     && 0.64      & 18.00     && 0.01       & 0.68      && 0.02       & 0.63      \\
SiS          & J=6--5                 & 108924.267                     && 1.10       & 30.00     && 1.86      & 52.40     && 0.05       & 2.44      && 0.04       & 1.14      \\
HC$_3$N      & J=12--11               & 109173.634                     && 2.03       & 52.39     && ---       & ---       && 0.07       & 3.16      && 0.08       & 2.79      \\
C$^{18}$O    & J=1--0                 & 109782.160                     && 0.02       & 0.76      && ---       & ---       && ---        & ---      && $<$\,0.01 & $<$\,0.17 \\
$^{13}$CO    & J=1--0                 & 110201.353                     && 1.69       & 26.62     && 2.50      & 36.40     && 0.31       & 14.93     && 0.06       & 1.82      \\
CH$_3$CN     & 6(1)--5(1)             & 110381.404                     && 0.09       & 4.77      && $<$\,0.05 & $<$\,1.42 && $<$\,0.01  & $<$\,0.18 && $<$\,0.01  & $<$\,0.23 \\
C$_2$S       & 8(9)--7(8)             & 113410.207                     && $<$\,0.03  & $<$\,0.48 && $<$\,0.07 & $<$\,1.96 && $<$\,0.02  & $<$\,0.53 && $<$\,0.01 & $<$\,0.11 \\
CN           & N=1--0                 & 113490.982                     && 3.43       & 238.43    && 3.88      & 310.20    && 0.13       & 16.43     && 0.19       & 14.71     \\
CO           & J=1--0                 & 115271.204                     && 10.29      & 269.43    && 20.83     & 542.71    && 1.27       & 55.86     && 1.11       & 36.71     \\
SiC$_2$      & 5$_{0,5}$--4$_{0,4}$   & 115382.38\phantom{0}           && 0.80       & 21.43     && 1.60      & 39.38     && 0.07       & 2.66      && 0.03       & 0.82      \\
C$^{18}$O    & J=2--1                 & 219560.319                     && 0.15       & 4.21      && ---       & ---       && ---        & ---       && ---        & ---       \\
$^{13}$CO    & J=2--1                 & 220398.686                     && 3.85       & 79.62     && ---       & ---       && 0.90       & 35.58     && 0.15       & 5.76      \\
CH$_3$CN     & 12(0)--11(0)           & 220747.268                     && 0.17       & 2.80      && ---       & ---       && $<$\,0.05  & $<$\,1.01 && $<$\,0.03  & $<$\,0.48 \\
CN           & N=2--1                 & 226874.564                     && 2.55       & 231.18    && ---       & ---       && 0.15       & 17.06     && 0.20       & 11.32     \\
CO           & J=2--1                 & 230538.000                     && 34.60      & 799.00    && ---       & ---       && 4.20       & 150.00    && 2.54       & 69.40     \\
CS           & J=5--4                 & 244935.606                     && 16.57      & 405.53    && ---       & ---       && 0.77       & 32.34     && 0.47       & 10.89     \\
HCN          & J=3--2                 & 265886.432                     && 45.10      & 1010.24   && ---       & ---       && 4.69       & 139.76    && 0.88       & 20.83     \\
\hline
\end{tabular}
\end{flushleft}
\end{table}
\clearpage
\begin{table}
\begin{flushleft}
\addtocounter{table}{-1}
\caption{(\emph{cont.}) Detected lines}
\little
\begin{tabular}{l l r r rr r rr r rr r rr}
\hline\hline
Molecule & Transition & Frequency &&\multicolumn{2}{c}{IRAS07454-7112}&&
\multicolumn{2}{c}{CIT\,6}&&\multicolumn{2}{c}{AFGL\,3068}&&\multicolumn{2}{c}{IRC+40540}\\
 \cline{5-6}\cline{8-9}\cline{11-12}\cline{14-15}
 & & && $T_{\rm mb}$ & $\int T_{\rm mb}{\rm d}v$ && $T_{\rm mb}$ & $\int T_{\rm mb}{\rm d}v$ && $T_{\rm mb}$ & $\int T_{\rm mb}{\rm d}v$ && $T_{\rm mb}$ & $\int T_{\rm mb}{\rm d}v$\\
 & & (MHz) && (K) & (K\,kms$^{-1}$)&& (K) & (K\,kms$^{-1}$)&& (K) & (K\,kms$^{-1}$) && (K) & (K\,kms$^{-1}$) \\
\hline
C$_2$S       & 6(7)--5(6)             & \phantom{0}86181.413           && $<$\,0.01  & $<$\,0.07 && $<$\,0.02  & $<$\,0.36 && $<$\,0.02 & $<$\,0.44 && $<$\,0.01  & $<$\,0.28 \\
H$^{13}$CN   & J=1--0                 & \phantom{0}86340.184           && 0.07       & 1.45      && 0.15       & 3.13      && 0.13      & 3.76      && 0.13       & 3.85      \\
C$_3$S       & J=15--14               & \phantom{0}86708.379           && $<$\,0.01  & $<$\,0.07 && $<$\,0.01  & $<$\,0.23 && $<$\,0.02 & $<$\,0.47 && $<$\,0.01  & $<$\,0.21 \\
SiO          & J=2--1            & \phantom{0}86846.998           && 0.03       & 0.52      && 0.17       & 3.16      && $<$\,0.02 & $<$\,0.47 && 0.04       & 0.94      \\
HN$^{13}$C   & J=1--0                 & \phantom{0}87090.859           && $<$\,0.01  & $<$\,0.08 && $<$\,0.01  & $<$\,0.23 && $<$\,0.02 & $<$\,0.47 && $<$\,0.01  & $<$\,0.25 \\
C$_2$H       & N=1--0                 & \phantom{0}87316.925           && $<$\,0.01  & $<$\,0.09 && 0.10       & 5.81      && 0.09      & 5.94      && $<$\,0.01  & $<$\,1.78 \\
HCN          & J=1--0                 & \phantom{0}88631.847           && 0.19       & 4.43      && 0.77       & 18.07     && 0.39      & 8.97      && 0.26       & 8.24      \\
C$_3$N       & N=9--8, J=19/2--17/2   & \phantom{0}89045.59\phantom{0} && $<$\,0.01  & $<$\,0.07 && ---        & ---       && ---       & ---       && ---        & ---       \\
C$_3$N       & N=9--8, J=17/2--15/2   & \phantom{0}89064.36\phantom{0} && $<$\,0.01  & $<$\,0.06 && ---        & ---       && ---       & ---       && ---        & ---       \\
HC$_5$N      & J=34--33               & \phantom{0}90525.892           && 0.02       & 0.06      && $<$\,0.01  & $<$\,0.21 && $<$\,0.01 & $<$\,0.21 && $<$\,0.02  & $<$\,0.37 \\
HC$^{13}$CCN & J=10--9                & \phantom{0}90593.059           && 0.01       & 0.13      && $<$\,0.01  & $<$\,0.11 && $<$\,0.01 & $<$\,0.10 && 0.08       & 0.35      \\
HCC$^{13}$CN & J=10--9                & \phantom{0}90601.791           && 0.01       & 0.13      && $<$\,0.01  & $<$\,0.11 && $<$\,0.01 & $<$\,0.10 && 0.08       & 0.35      \\
HNC          & J=1--0                 & \phantom{0}90663.543           && 0.02       & 0.50      && 0.13       & 2.95      && 0.06      & 1.18      && 0.09       & 1.42      \\
SiS          & J=5--4                 & \phantom{0}90771.546           && 0.03       & 1.07      && 0.04       & 0.59      && 0.04      & 0.59      && 0.15       & 4.11      \\
HC$_3$N      & J=10--9                & \phantom{0}90978.993           && 0.06       & 1.39      && 0.19       & 5.17      && 0.12      & 3.37      && 0.19       & 6.16      \\
$^{13}$CS    & J=2--1                 & \phantom{0}92494.299           && $<$\,0.01  & $<$\,0.04 && ---        & ---       && ---       & ---       && ---        & ---       \\
SiC$_2$      & 4$_{0,4}$--3$_{0,3}$   & \phantom{0}93063.639           && 0.01       & 0.13      && ---        & ---       && ---       & ---       && ---        & ---       \\
HC$_5$N      & J=35--34               & \phantom{0}93188.127           && $<$\,0.01  & $<$\,0.04 && ---        & ---       && ---       & ---       && ---        & ---       \\
C$_4$H       & N=10--9, J=21/2--19/2  & \phantom{0}95150.32\phantom{0} && $<$\,0.01  & $<$\,0.03 && $<$\,0.01  & $<$\,0.13 && ---       & ---       && ---        & ---       \\
C$_4$H       & N=10--9, J=19/2--17/2  & \phantom{0}95188.94\phantom{0} && $<$\,0.01  & $<$\,0.03 && $<$\,0.01  & $<$\,0.13 && ---       & ---       && ---        & ---       \\
SiC$_2$      & 4$_{2,2}$--3$_{2,1}$   & \phantom{0}95579.389           && $<$\,0.01  & $<$\,0.06 && ---        & ---       && ---       & ---       && ---        & ---       \\
CS           & J=2--1                 & \phantom{0}97980.968           && 0.13       & 2.53      && 0.67       & 16.91     && 0.14      & 3.93      && 0.30       & 9.56      \\
C$_3$H       & $^2\Pi_{1/2}$,9/2--7/2 & \phantom{0}97995.450           && $<$\,0.01  & $<$\,0.09 && $<$\,0.02  & $<$\,0.55 && $<$\,0.01 & $<$\,0.31 && $<$\,0.02  & $<$\,0.45 \\
HC$^{13}$CCN & J=12--11               & 108710.523                     && 0.01       & 0.08      && 0.05       & 0.44      && $<$\,0.01 & $<$\,0.14 && $<$\,0.01  & $<$\,0.09 \\
HCC$^{13}$CN & J=12--11               & 108721.008                     && 0.01       & 0.08      && 0.05       & 0.44      && $<$\,0.01 & $<$\,0.14 && $<$\,0.01  & $<$\,0.09 \\
$^{13}$CN    & N=1--0                 & 108780.201                     && 0.01       & 0.73      && 0.05       & 4.21      && $<$\,0.01 & $<$\,0.28 && $<$\,0.01  & $<$\,0.18 \\
C$_3$N       & N=11--10, J=23/2--21/2 & 108834.27\phantom{0}           && 0.01       & 0.23      && 0.10       & 2.19      && 0.05      & 1.34      && 0.03       & 0.70      \\
C$_3$N       & N=11--10, J=21/2--19/2 & 108853.02\phantom{0}           && 0.01       & 0.23      && 0.10       & 2.19      && 0.05      & 1.34      && 0.03       & 0.70      \\
SiS          & J=6--5                 & 108924.267                     && 0.03       & 0.88      && 0.19       & 3.62      && 0.15      & 2.23      && 0.07       & 1.91      \\
HC$_3$N      & J=12--11               & 109173.634                     && 0.06       & 1.48      && ---        & ---       && ---       & ---       && ---        & ---       \\
$^{13}$CO    & J=1--0                 & 110201.353                     && 0.09       & 2.10      && 0.40       & 5.90      && 0.17      & 9.33      && 0.22       & 4.39      \\
CH$_3$CN     & 6(1)--5(1)             & 110381.404                     && $<$\,0.01  & $<$\,0.12 && $<$\,0.01  & $<$\,0.29 && $<$\,0.03 & $<$\,0.83 && $<$\,0.03  & $<$\,0.72 \\
C$_2$S       & 8(9)--7(8)             & 113410.207                     && $<$\,0.01  & $<$\,0.10 && $<$\,0.03  & $<$\,0.89 && $<$\,0.02 & $<$\,0.53 && $<$\,0.02  & $<$\,0.64 \\
CN           & N=1--0                 & 113490.982                     && 0.17       & 10.30     && 1.12       & 61.84     && 0.09      & 5.97      && 0.41       & 21.84     \\
CO           & J=1--0                 & 115271.204                     && 1.16       & 25.57     && 3.52       & 105.00    && 2.08      & 47.08     && 1.94       & 41.88     \\
SiC$_2$      & 5$_{0,5}$--4$_{0,4}$   & 115382.38\phantom{0}           && $<$\,0.01  & $<$\,0.24 && 0.23       & 6.71      && $<$\,0.02 & $<$\,0.62 && $<$\,0.03  & $<$\,0.75 \\
$^{13}$CO    & J=2--1                 & 220398.686                     && 0.25       & 7.34      && ---        & ---       && ---       & ---       && ---        & ---       \\
CH$_3$CN     & 12(0)--11(0)           & 220747.268                     && 0.04       & 0.67      && ---        & ---       && ---       & ---       && ---        & ---       \\
CN           & N=2--1                 & 226874.564                     && 0.41       & 20.59     && ---        & ---       && ---       & ---       && ---        & ---       \\
CO           & J=2--1                 & 230538.000                     && 2.60       & 50.00     && ---        & ---       && ---       & ---       && ---        & ---       \\
CS           & J=5--4                 & 244935.606                     && 0.49       & 8.12      && ---        & ---       && ---       & ---       && ---        & ---       \\
HCN          & J=3--2                 & 265886.432                     && 0.50       & 8.82      && ---        & ---       && ---       & ---       && ---        & ---       \\
\hline
\end{tabular}
\end{flushleft}
\end{table}
\normalsize
\end{landscape}

\subsection{Upper limits}

The integrated intensities for lines which were not detected are
determined using,
\begin{eqnarray}
I_v \leq 3\sigma \left(\sqrt{2}\ \sqrt{2\frac{v_{\mathrm{exp}}}{\Delta v_{\mathrm{res}}}}\right)\Delta\nu = 3\sigma
(2n)^{1/2}\Delta\nu.
\end{eqnarray}
where $\sigma$\ is the rms noise in the spectrum, $\Delta\nu$\ is the
frequency resolution of the spectrum, $v_{\mathrm{res}}$\ the velocity
resolution of the spectrum and $n$\ the number of channels covering
the line width. Integrated intensities calculated with this method
were used to determine upper limits to abundances.

\section{Full radiative transfer modelling}
\label{modelling}
In order to obtain accurate values of the mass-loss rates for the
objects in the sample, detailed non-LTE radiative transfer modelling
of the CO line emission was carried out.  This modelling has proven to
be one of the most reliable methods for estimating many of the
parameters characterising a circumstellar envelope, in particular, the
mass-loss rate, kinetic temperature structure, and expansion velocity
\citep{Kastner1992, Groenewegen94, Crosas97, SchoierOlofsson2001,
Schoier_etal2002, Olofsson_etal2002}.  Once the basic envelope
parameters are known (as well as the stellar parameters), the
abundances of various molecules present in the circumstellar envelopes
can be determined.

\begin{table*}
\begin{flushleft}
\caption{Summary of circumstellar properties derived from the CO modelling (see text for details)}
\label{tabmlr}
\begin{tabular}{lccccccccc}
\hline\hline
Source &  $\dot{M}$ &  $h$ & $N$$^b$ & $\chi^2_{\mathrm{red}}$$^c$ & $v_{\rm exp}$ &  $(\dot{M}/v_{\rm exp})_{\star}$ & $r_{\rm p}^d$ &  $d$ & $^{12}$CO$^e$\\
\cline{7-7}
\cline{10-10}
 & [M$_{\odot}$\,yr$^{-1}$] & & & & [${\rm km\,s}^{-1}$] & $(\dot{M}/v_{\rm exp})_{10216}$ & [cm] &  [cm] & $^{13}$CO  \\
\hline
IRAS07454-7112 &  5.0\,(-6) &1.5$^a$                     & \phantom{0}2   & 0.1 &13.0 & 0.46  & 3.0\,(17)   & 2.4\,(16)  & 17 \\
IRC+10216            & 1.2\,(-5) & 1.5\phantom{$^a$} & 27                       & 1.4 & 14.5 & 1.00 & 3.7\,(17)   & 5.5\,(16) & 45        \\
CIT\,6                      &  5.0\,(-6) & 2.3\phantom{$^a$} & 20                      & 0.8 & 17.0 & 0.36 & 1.9\,(17)   & 2.8\,(16) & 35     \\
IRAS15082-4808 &  1.0\,(-5) & 1.5$^a$                    & \phantom{0}2   & 0.2 & 19.5 & 0.62 & 2.5\,(17)    & 3.5\,(16) & 35          \\
IRAS15194-5115 &  1.2\,(-5) & 2.8\phantom{$^a$}& 10                       & 0.9& 21.5 & 0.67 & 3.2\,(17)     & 7.2\,(16) & \phantom{0}6      \\
AFGL\,3068           &  2.0\,(-5) & 2.5\phantom{$^a$} &  \phantom{0}4 & 1.6& 14.0 & 1.72 & 3.8\,(17)   & 1.2\,(17) & 30        \\
IRC+40540           &  1.5\,(-5) & 1.5\phantom{$^a$} & \phantom{0}7   & 1.0 & 14.0 & 1.29 & 4.0\,(17)  & 7.3\,(16) & 50  \\
\hline
\end{tabular}
  {\footnotesize
   \begin{enumerate}
 	\renewcommand{\labelenumi}{(\alph{enumi})}
         \item Adopted value
          \item Number of observational constraints used in the $^{12}$CO modelling
         \item Reduced $\chi^2$ of the best fit $^{12}$CO model
	   \item $r_{\rm p}$\ is the photodissociation radius of CO
         \item Determined from radiative transfer modelling of both $^{12}$CO and $^{13}$CO emission
   \end{enumerate}
     }
\end{flushleft}
\end{table*}

\subsection{CO line modelling}
Here the detailed non-LTE radiative transfer code of
\citet{Schoier2000} is adopted and the modelling procedure as outlined
in detail in \citet{SchoierOlofsson2001} is used.  The code has been
tested against a wide variety of molecular line radiative transfer
codes, for a number of benchmark problems, to a high accuracy
\citep{Zadelhoff_etal2002}.  The Monte Carlo method \citep{Bernes1979}
is used to determine the steady-state level populations of the CO
molecules in the envelope as a function of distance from the star,
using the statistical equilibrium equations.  In addition, the code
simultaneously solves the energy balance equation including the most
relevant heating and cooling processes. Heating is dominated by
collisions between the dust and gas except in the outermost parts of
the envelope where the photoelectric effect effectively heats the
envelope. Cooling is generally dominated by molecular line cooling
from CO but adiabatic cooling due to the expansion of the envelope is
also important. The excitation analysis allows for a self-consistent
treatment of CO line cooling.

The observed circumstellar CO line emission is modelled taking into
account 50 rotational levels in each of the fundamental and first
excited vibrational states.  The energy levels and radiative
transition probabilities from \citet{Chandra_etal1996} are used.  The
recently published collisional rates of CO with H$_2$ by
\citet{Flower2001} have been adopted assuming an ortho-to-para ratio
of 3.  For temperatures above 400\,K the rates from
\citet{Schinke_etal1985} were used and further extrapolated to include
transitions up to $J$=50.  The collisional rates adopted here differ
from those used in the previous modelling of some of these sources
\citep{Ryde_etal1999, SchoierOlofsson2000, SchoierOlofsson2001} and
account for the slightly different envelope parameters derived in the
present analysis. The same set of collisional rates were used for all
CO isotopomers.

The envelopes are assumed to be spherically symmetric and to expand at
a constant velocity and the model includes the radiation emitted from
the central star. Dust present around the star will absorb parts of
the stellar radiation and re-emit it at longer wavelenghts. For
simplicity, the central radiation field is represented by one or two
blackbodies and is determined from a fit to the observed spectral
energy distribution (SED) (Table~\ref{stellardata}).  The inner radius
of the circumstellar envelope is taken to reside outside that of the
central blackbodies. This procedure provides a good description of the
radiation field to which the envelope is subjected.  For the sample
stars, which have dense CSEs, the line intensities derived from the
$^{12}$CO model are not sensitive to the adopted description of the
stellar spectrum due to the high line optical depths.  The stellar
photons are typically absorbed within the first few shells in the
model. A more detailed treatment of thermal emission from the dust
present in the CSE, and the increase of total optical depth at the
line wavelenghts, has been found to be of no major importance in
deriving the envelope parameters for high
mass-loss rate objects \citep{Schoier_etal2002}.

The abundance of $^{12}$CO relative to H$_2$ was fixed at $1.0\times
10^{-3}$, in agreement with \citet{Willacy98} and the survey of
\citet{Olofsson_etal1993b}. The CO envelope size was estimated based
upon modelling results from \citet{Mamon88}, which have been shown to
compare well with observations \citep{SchoierOlofsson2001}.  The same
envelope size was assumed for all CO isotopomers.

\subsection{Fitting strategy}
With the assumptions made in the standard model there remains two
principal free parameters, the mass-loss rate ($\dot{M}$) and the
so-called $h$-parameter, which contains information about individual
dust grains and controls the amount of heating through dust-gas
collisions.
\begin{equation}
\label{h-para}
h = \left( \frac{\Psi}{0.01} \right)
    \left(\frac{2.0\,\mathrm{g\,cm}^{-3}}{\rho_{\mathrm{d}}} \right)
    \left(\frac{0.05\,{\mathrm{\mu m}}}{a_{\mathrm{d}}} \right), 
\end{equation}
where $\Psi$ is the dust-to-gas mass-loss rate ratio, 
$\rho_{\mathrm{d}}$ the mass density of a dust grain, and
$a_{\mathrm{d}}$ the size of a dust grain, and the adopted values of
these parameters are also given.  The mass-loss rate and the
$h$-parameter were allowed to vary simultaneously.  Once the molecular
excitation, i.e., the level populations, is obtained the radiative
transfer equation can be solved exactly.  The resulting brightness
distribution is then convolved with the appropriate beam to allow a
direct comparison with observations.  The beam profile used in the
convolution of the modelled emission is assumed to be Gaussian which
is appropriate at the frequencies used here.  The best fit
circumstellar model for a particular source is estimated from a
$\chi^2$-analysis (for details see \citealt{SchoierOlofsson2001})
using observed $^{12}$CO integrated intensities and assuming a
15--20\% calibration uncertainty.  The number of observational
constraints used in the modelling is shown in Table~\ref{tabmlr}
together with the reduced $\chi^2$ of the best fit model.  In all the
cases $\chi^2_{\mathrm{red}}$$\sim$1, indicating good fits.

\begin{figure*}
   \centering{   
   \includegraphics[width=18cm]{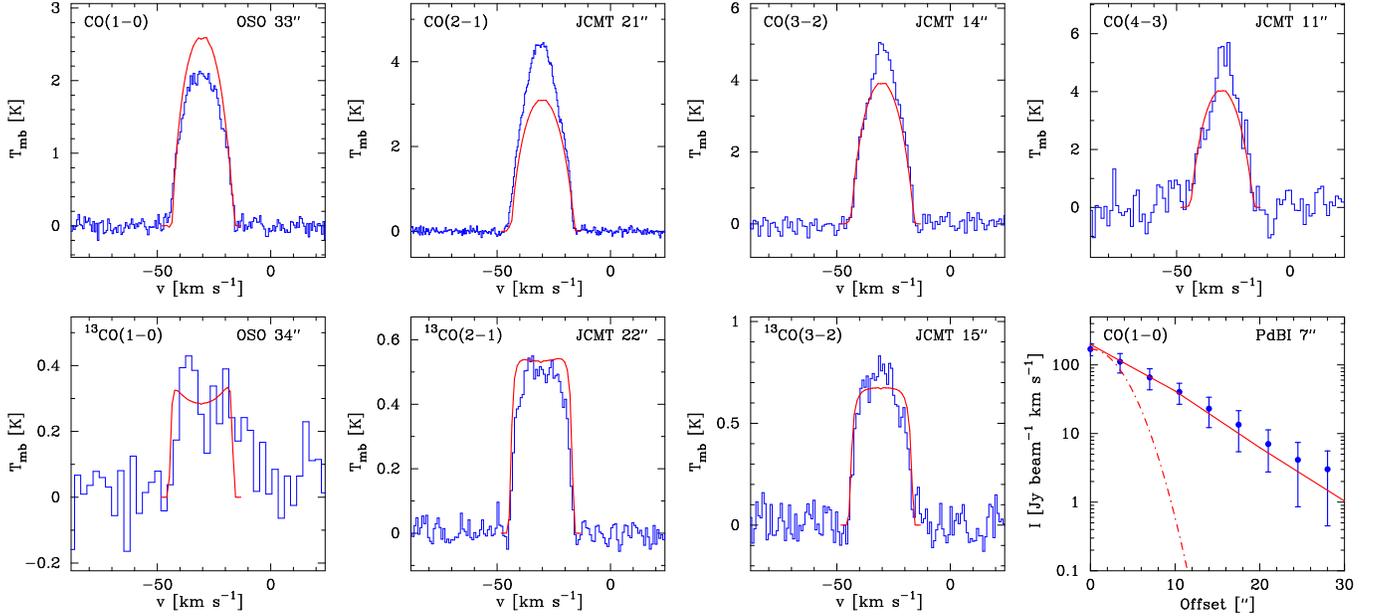}
   \caption{Multi-transition CO millimetre-wave line emission observed
   towards \object{AFGL 3068}. The observed spectra (histograms) have
   been overlayed with the best fit model results (solid lines). Also
   shown, lower right panel, is the observed radial brightness
   distribution (boxes with error bars) overlayed by the results from
   the model (full line), with the circular beam used in the radiative
   transfer analysis (dot-dashed line). The transition, telescope
   used, and the corresponding beamsize, are indicated for each
   observation.}
   \label{afgl}}
   \end{figure*}

The $^{12}$CO data used in the analysis are presented in
\citet{SchoierOlofsson2001} and \citet{Schoier_etal2002} for
\object{CIT 6}, \object{IRC+10216}, \object{IRAS 15194--5115}, and
\object{IRC+40540} and consist of both millimetre and sub-millimetre
line data as well as far infra-red high-$J$ transitions observed by
ISO.  The best fit $^{12}$CO model obtained for \object{AFGL 3068} is
overlayed onto observations and presented in Fig.~\ref{afgl}. For
\object{IRAS 07454--7112} and \object{IRAS 15082--4808} only $J$=1--0
and $J$=2--1 line data as observed by the SEST, and presented here in
Figs.~\ref{07454lrs1}, \ref{07454lrs2}, \ref{07454hrs1},
\ref{07454hrs2} and \ref{15082hrs1}--\ref{15082lrs2}, are available.
Due to the limited number of constraints the $h$-parameter was assumed
to be equal to 1.5, i.e., the dust properties were taken to be the
same as for \object{IRC+10216} and \object{IRC+40540} for these two
sources.

\subsection{The mass-loss rates}
The determination of accurate mass-loss rates for some of these carbon
stars is, it turns out, difficult in models where CO cooling is
treated in a self-consistent manner \citep{Sahai1990, Kastner1992,
SchoierOlofsson2001}.  The reason for this is that an increase in
$\dot{M}$ leads to more net cooling than heating, which compensates
for the increase in molecular density, making it hard to
simultaneously constrain both $\dot{M}$ and the $h$-parameter.  For
the sources \object{CIT 6}, \object{IRC+10216} and \object{IRAS
15194--5115}, where a significant number of high-$J$ transitions were
observed by ISO, the degeneracy between $\dot{M}$ and the
$h$-parameter is partly lifted allowing for better constraints to be
put on these parameters \citep{Schoier_etal2002}.  For these three
sources the mass-loss rate is estimated to be accurate to about 50\%
within the adopted circumstellar model.  For the remaining sources the
uncertainty in the derived mass-loss rate is larger due to either the
above mentioned degeneracy (\object{AFGL 3068} and \object{IRC+40540})
or that only two $^{12}$CO lines are used in the analysis
(\object{IRAS 07454--7112} and \object{IRAS 15082--4808}). Radiative
transfer analysis of the observed continuum emission from these
sources (Sch\"oier et~al., in prep) give mass-loss rates that agree
within $\sim$50\% when compared with those derived from the CO
modelling. This is reassuring and lends further credibility to the
mass-loss rates presented in Table~\ref{tabmlr}.

The CSEs of the sample stars have similar physical
properties. However, the stars presented here are losing mass at a
significantly higher rate than the average carbon star:
\citet{SchoierOlofsson2001} measure a median mass-loss rate for carbon
stars of $\sim$3$\times$10$^{-7}$\,M$_{\odot}$\,yr$^{-1}$, based on a
sample of carbon stars complete within $\sim 600$\,pc from the
sun. This suggests that the stars in the sample presented here are
going through the super-wind phase of evolution
\citep[e.g.][]{VassWood1993} at the end of the AGB, and will soon
eject the entire stellar mantle.

\subsection{CO isotopic ratios}
Using the envelope parameters derived from the $^{12}$CO modelling,
the abundance of $^{13}$CO and C$^{18}$O was estimated using the same
radiative transfer code.  The derived $^{12}$CO/$^{13}$CO-ratios are
presented in Table~\ref{tabmlr}.  The best fit $^{13}$CO model
obtained for \object{AFGL 3068} is overlayed onto observations and
presented in Fig.~\ref{afgl}.  The fits to the observed $^{13}$CO
emission are good in all cases with
$\chi^2_{\mathrm{red}}$$\lesssim$1. The quality of the fits further
strengthen the adopted physical models for the envelopes.  The
$^{12}$CO/$^{13}$CO-ratio derived in this manner allows the assumption
of optically thin emission adopted in Sect.~\ref{abundances} to be
tested for those molecular species where the isotopomer containing
$^{13}$C has been detected.  All the sample stars, except for
\object{IRAS 15194--5115} (to be discussed in Sect.~\ref{iras15194}),
have inferred $^{12}$C/$^{13}$C-ratios in the range $\sim 20-50$.  For
optically bright carbon stars, i.e. generally lower mass-loss rate
objects, values between $\sim 20-90$ are found
\citep[e.g.,][]{SchoierOlofsson2000, Abia01}.  The observed
$^{12}$C/$^{13}$C-ratios are in agreement with evolutionary scenarios
where the $^{12}$C/$^{13}$C-ratio is thought to increase from an
initial low value of $\sim 10-25$, that depends on stellar mass, as
the star evolves along the AGB \citep{Abia01}. The thermal pulses that
an AGB star experiences will effectively dredge nucleosynthesized
$^{12}$C to the surface \citep{Iben83} and, in addition to increasing
the $^{12}$C/$^{13}$C-ratio, eventually turn it into a carbon star
with C/O$>$$1$ in its atmosphere.

C$^{18}$O emission was only detected towards \object{IRC+10216} and a
C$^{16}$O/C$^{18}$O-ratio of 1050 is derived, using four observational
constraints (including three different transitions $J$=1--0, 2--1,
3--2) for C$^{18}$O.  This value is in excellent agreement with the
$^{16}$O/$^{18}$O-ratio of 1260 obtained by \citet{Kahane92}, using a
combination of optically thin emission lines.  In comparison, the
value of this ratio in the solar neighbourhood is around 500.  The
$^{16}$O/$^{18}$O-ratio and in particular the $^{17}$O/$^{18}$O-ratio,
which is not measured here, can be used as tracers of nucleosynthesis.
Like the $^{12}$C/$^{13}$C-ratio, these ratios are thought to increase
as the star evolves along the AGB. \citet{Kahane92} indeed find
support for this scenario in a small sample of carbon-rich evolved
stars.

\section{Abundance calculations}
\label{abundances}
A full radiative transfer analysis of the wealth of molecular data
presented here is beyond the scope of this paper. A detailed treatment
of the excitation would have to include the effects of dust emission
and absorption in addition to accurate rates for collisional
excitation of the molecules.  Instead, abundances have been calculated
for species with emission that is expected to be optically thin. In
the case that the emission is optically thick, the calculated
abundances using this assumption will be lower limits.

The calculation of isotope ratios of various species (as shown in
Table~\ref{12c/13c}) shows in general that many lines are optically
thick (i.e., the ratio derived from observations is lower than the
$^{12}$CO/$^{13}$CO abundance ratio derived from the radiative
transfer analysis). Where this is the case, the abundance of the main
isotope (viz., CN, CS and HC$_3$N) has been calculated by scaling the
abundance of the less abundant isotope by the
$^{12}$CO/$^{13}$CO-ratio. This is indicated by the bold-faced type in
Table~\ref{tababsiz}. The abundance of HCN is not calculated using
Eq.~\ref{optthin} since the line is certainly optically thick, but in
all cases only by scaling the abundance of H$^{13}$CN by the
calculated $^{12}$CO/$^{13}$CO-ratio.

\subsection{Radiative transfer}
In the calculation of molecular abundances the circumstellar envelope
is assumed to have been formed by a constant mass-loss rate and to
expand with a constant velocity, such that the total density
distribution follows an $r^{-2}$\ law. It is further assumed that the
fractional abundance of a species is constant in the radial range
$r_{\mathrm{i}}$\ to $r_{\mathrm{e}}$\ and zero outside it. The
excitation temperature was assumed to be constant throughout the
emitting region.  With these assumptions \citet{Olofsson_etal1990}
showed that for a given molecular transition,
\begin{eqnarray}
\label{optthin}
\int T_{\mathrm{mb}}{\mathrm{d}}v
&&= \frac{c^2}{2k\nu^2}\left[ B_{\nu}(T_{\mathrm{ex}}) - B_{\nu}(T_{\mathrm{bg}}) \right] \times \nonumber\\
&&\times\ \frac{g_uA_{ul}}{8\pi}\frac{c^3}{\nu^3}\frac{(1-e^{-h\nu/kT_{\mathrm{ex}}})}{Z}e^{-E_l/kT_{\mathrm{ex}}}\times
\nonumber\\
&& \times\ \frac{f_{\mathrm{X}}\dot{M}_{\mathrm{H_2}}}{v_{\mathrm{exp}}m_{\mathrm{H_2}}
BD}\int_{x_{\mathrm{i}}}^{x_{\mathrm{e}}}e^{-4{\ln(2)}x^2}{\mathrm{d}} x,
\end{eqnarray}
where $c$ is the speed of light, $h$ the Planck constant, $k$ the
Boltzmann constant, $B_{\nu}$ the Planck function, $T_{\mathrm{ex}}$
the excitation temperature, $T_{\mathrm{bg}}$ is taken to be the
blackbody temperature of the cosmic background radiation at 2.7\,K,
$m_{\mathrm{H_2}}$\ is the mass of an H$_2$\ molecule,
$v_{\mathrm{exp}}$ the expansion velocity of the circumstellar
envelope, $B$ the FWHM of the telescope beam, $D$ the assumed
distance, $\dot{M}_{\mathrm{H_2}}$ the mass-loss rate, $Z$ the
partition function, $\nu$ the frequency of the line, $g_{\mathrm{u}}$
the statistical weight of the upper level, $A_{\mathrm{ul}}$ the
Einstein coefficient for the transition, $E_{\mathrm{l}}$ the energy
of the lower transition level, and
$x_{\mathrm{i,e}}=r_{\mathrm{i,e}}/BD$.  The integral over $x$ takes
care of the beam filling.  However, if $r\ll BD$, the integral can be
simplified to $(r_{\mathrm{e}}-r_{\mathrm{i}})/BD$. This is the case
in the furthest sources, \object{IRAS 07454--7112}, \object{CIT 6} and
\object{IRAS 15082--4808}.  However, for the CN emission from these
sources, which tends to be very extended, and for the remaining four
sources, the full integral is calculated.

   \begin{figure}
   \centering{   
   \includegraphics[width=8.8cm]{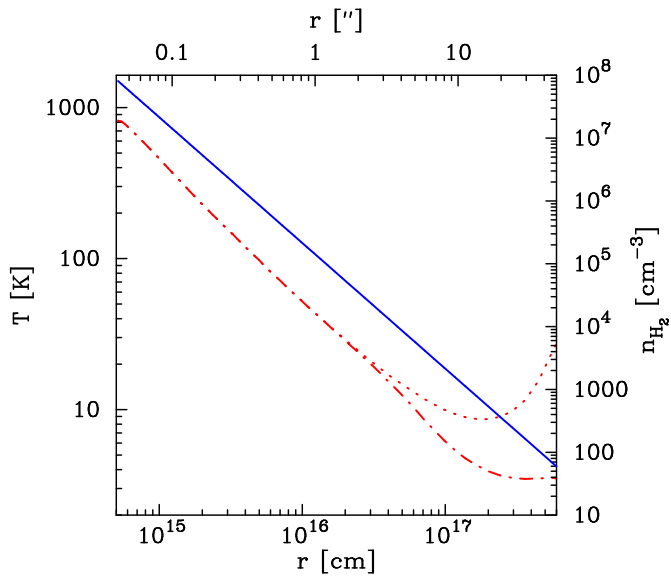}
   \caption{The density (solid line) and kinetic temperature (dotted line) structures obtained from the CO excitation analysis
for \object{AFGL 3068}. Also
shown is the excitation temperature of the CO($J$=2--1) line (dash-dotted line).}
   \label{afgl_struct}}
   \end{figure}

\begin{table*}
\begin{flushleft}
\caption{Rotation temperatures}
\label{tabrotemp}
\begin{tabular}{l cccccccccccc}
\hline\hline
Source && $^{13}$CO && CN && CS && C$_3$N && C$_4$H && HCN \\
 && 2--1/1--0 && 2--1/1--0 && 5--4/2--1 && 11--10/10--9 && 10--9/9--8 && 3--2/2--1\\ 
\hline
\object{IRAS 07454--7112}     && 7.0 && 5.2 && 11.3           && ---  && --- && 5.7 \\
\object{IRC+10216} (S)        && 6.3 && 3.9 && 13.7           && 17.9 && 5.3 && 7.8 \\
\object{IRAS15082--4808}      && 6.5 && 3.6 && \phantom{0}8.3 && ---  && --- && 5.6 \\
\object{IRAS15194--5115}      && 5.6 && 4.0 && \phantom{0}9.7 && ---  && --- && 9.7 \\
Average       && 6.4 && 4.2 && 10.8           && 17.9 && 5.3 && 7.2 \\
\hline
\qquad\\
\hline
Source && H$^{(13)}$C$_3$N && HC$_3$N && HC$_5$N && SiC$_2$ && SiS\\
 && 12--11/10--9 && 12--11/10--9 && 34--33/32--31 && 5-4/4-3 && 6--5/5--4 && Average \\
\hline
\object{IRAS07454--7112}     && \phantom{0}8.5 && 15.2           && ---  && ---  && \phantom{0}5.7 && \phantom{0}8.4 \\
\object{IRC+10216} (S)       && 17.8           && 10.7           && 10.0 && 11.6 && \phantom{0}5.8 && 10.1 \\ 
\object{IRC+10216} (O)       && ---            && ---            && ---  && ---  && 11.8           && 11.8 \\
\object{IRAS15082--4808}     && ---            && \phantom{0}8.9 && ---  && 37.1 && \phantom{0}5.0 && 10.7 \\ 
\object{IRAS15194--5115}     && \phantom{0}7.2 && 11.3           && ---  && ---  && \phantom{0}6.1 && \phantom{0}7.7 \\ 
\object{IRC+40540}           && ---            && ---            && ---  && ---  && \phantom{0}3.5 && \phantom{0}3.5 \\
\cline{13-13}
Average                      && 11.2           && 11.5           && 10.0 && 24.4 && \phantom{0}6.3 && {\bf\phantom{0}8.7} \\
\hline
\end{tabular}
\end{flushleft}
\end{table*}

\subsection{The excitation temperature}
Of importance in the radiative transfer analysis is the excitation
(rotational) temperature assumed for the molecular emission.  In
Fig.~\ref{afgl_struct} the kinetic temperature and density structures
for \object{AFGL 3068} are shown. In the region where most of the
molecular emission observed is thought to emanate from,
\hbox{$\sim10^{16}-10^{17}$\,cm,} the kinetic temperature ranges from
\hbox{$\sim30-10$\,K.} Thus, if the emission were to be fully
thermalized, then excitation temperatures would be expected to lie
within this range.  However, in addition to the low temperatures, the
relatively low densities in this region,
\hbox{$\sim10^2-10^5$\,cm$^{-3}$,} makes the CO emission partly
sub-thermally excited as illustrated in Fig.~\ref{afgl_struct}. Thus,
sub-thermal emission is to be expected for all molecular species and
the excitation temperature will depend on the species (and transition)
observed.

When two or more transitions of the same molecule are observed, it is
possible to make an estimate of the rotation temperature
($T_{\mathrm{rot}}$) of that molecule using
Eq.~\ref{optthin}. Assuming a molecular species to be excited over the
same radial range, and according to a single temperature, the
rotational temperature can be estimated. The results are shown in
Table \ref{tabrotemp}. It is clear that the rotational temperatures
vary from source to source and between molecular species in the range
\hbox{$\sim3-30$\,K,} as to be expected. The average excitation temperature
is 8.7\,K (averaged over the individual excitation temperatures for
all sources, rather than molecular species). A generic value of 10\,K
was assumed for all the abundances estimated.

\subsection{The partition function}

All molecules are assumed to be linear, rigid rotators, except for
SiC$_2$ and C$_3$H$_2$\ which are asymmetric tops and CH$_3$CN, which
is a prolate symmetric top. Einstein A-coefficients (where available)
and energy levels are taken from \citet{ChandraRashmi1998} for
SiC$_2$\ , from \citet{Vrtilek_etal1987} for C$_3$H$_2$, and from
\citet{Boucher_etal1980} for CH$_3$CN.

The partition function, $Z$, is calculated assuming that no molecules
have a hyperfine structure, i.e. as having simple rotational energy
diagrams. Hence the integrated intensities which are summed over all
hyperfine components are used. For C$_3$H$_2$, CH$_3$CN and SiC$_2$\
we use the approximate expression for an asymmetric rotor
\citep{TownesSchawlow1975} multiplied by 2 for C$_3$H$_2$, by 4 for
CH$_3$CN (to account for spin statistics) and by 1/2 for SiC$_2$\
(since half of the energy levels are missing because of spin
statistics). It is assumed that all levels are populated according to
$T_{\mathrm{ex}}$.

\subsection{Sizes of emission regions}
\label{sizes}
\subsubsection{Photodissociation radii}
The chemical richness observed towards carbon stars can be
qualitatively understood in terms of a photodissociation model
\citep[see the review by][]{Glassgold1996}. Carbon-bearing molecules
like CO, C$_2$H$_2$, HCN and CS, in addition to Si-bearing molecules
like SiS and SiO, are all thought to be of photospheric origin where
they are formed in conditions close to LTE.  The photodissociation of
these so-called parent species produces various radicals and ions that,
in turn, drive a complex chemistry through ion-molecule and
neutral-neutral reactions. For example, the radicals C$_2$H and CN are
the photodissociation products of C$_2$H$_2$ and HCN, respectively.
In addition, the formation of dust in the CSE will affect the
abundances of some of the species, in particular SiO and SiS, which
condense onto dust grains.  All other species observed in the sample
are thought to be products of the circumstellar chemistry.

\begin{table}
\begin{flushleft}
\caption{Photodissociation rates}
\label{tabphdist}
\begin{tabular}{lc}
\hline\hline
Molecule &$G_0$       \\
         &[s$^{-1}$]    \\
\hline
HCN     &       1.1\,10$^{-9\phantom{0}}$\\
CN      &       2.5\,10$^{-10}$\\
C$_2$H$_2$&     2.1\,10$^{-9\phantom{0}}$\\
C$_2$H  &       3.4\,10$^{-10}$ \\
SiS     &       6.3\,10$^{-10}$ \\
CS      &       6.3\,10$^{-10}$ \\
SiO     &       6.3\,10$^{-10}$ \\
\hline
\end{tabular}
\end{flushleft}
\end{table}

To calculate the radial extent of the molecules HCN, CN, C$_2$H and CS
the photodissociation model of \citet{HugginsGlass1982} is
adopted. The photodissociation radius of a parent species
($f_{\mathrm{X}}$, viz., HCN, C$_2$H$_2$, CS) is determined by
\begin{equation}
\frac{{\mathrm{d}} f_{\mathrm{X}}}{{\mathrm{d}}
r}=-\frac{G_{\mathrm{0,X}}}{v_{\mathrm{exp}}}{\mathrm{e}}^{-d_{\mathrm{X}}/r}f_{\mathrm{X}}
\end{equation}
and of a daughter ($f_{\mathrm{Xd}}$, viz., C$_2$H) by
\begin{equation}
\frac{{\mathrm{d}} f_{\mathrm{Xd}}}{{\mathrm{d}}
r}=\frac{G_{\mathrm{0,X}}}{v_{\mathrm{exp}}}{\mathrm{e}}^{-d_{\mathrm{X}}/r}f_{\mathrm{X}}-\frac{G_{\mathrm{0,Xd}}}{v_{\mathrm{exp}}}{\mathrm{e}}^{-d_{\mathrm{Xd}}/r}f_{\mathrm{Xd}}
\end{equation}
$G_0$\ is the unshielded photodissociation rate and $d_{\mathrm{X}}$\
is the dust shielding distance, given by:
\begin{equation}
d_{\mathrm{X}} = 1.4 \frac{3 Q_{\mathrm{X}}}{4 a_{\mathrm{d}} \rho_{\mathrm{d}}} \frac{\dot{M}_{\mathrm{d}}}{4\pi
v_{\mathrm{d}}}, \label{sdist1}
\end{equation}
where $Q$ is the dust absorption efficiency, $\dot{M}_{\mathrm{d}}$
the dust mass-loss rate, and $v_{\mathrm{d}}$ the expansion velocity
of the dust grains \citep{JuraMorris1981}. Since different molecules
are generally dissociated at different wavelengths the adopted value
of $Q$ will depend on the species under study. However, the wavelength
dependence of $Q$ in the region of interest, $\sim$1000--3000\,{\AA},
is weak \citep[e.g.,][]{Suh2000} and a generic value of 1.0 for most
species of interest here is adopted.  However, the shielding distance
of CN is taken to be a factor of 1.2 greater than that used for HCN
\citep{Truong_etal1987}.  Introducing the $h$-parameter defined in
Eq.~\ref{h-para} the dust shielding distance may now be expressed as
\begin{equation}
d = 5.27\,10^{22} \, \frac{h \dot{M}}{v_{\mathrm{d}}} \quad{\mathrm{cm}}, \label{sdist2}
\end{equation}
where the H$_2$ mass-loss rate $\dot{M}$\ is given in
M$_{\odot}$\,yr$^{-1}$\ and $v_{\mathrm{d}}$\ in ${\mathrm{km\,s}}^{-1}$.
A gas-to-dust mass ratio of 0.01 was assumed.

%
%
There will generally be a drift velocity between the dust and the gas
 \citep[e.g.,][]{SchoierOlofsson2001}
\begin{equation}
v_{\mathrm{d}} - v_{\mathrm{exp}} = \sqrt{\frac{L v_{\mathrm{exp}} \langle Q\rangle}{\dot{M} c}},
\end{equation}
where $L$ is the bolometric luminosity of the star, $c$ the speed of
light, and $\langle Q\rangle$ is the flux-averaged momentum transfer
efficiency. The drift velocity also enters in the expression of the
heating caused by dust-gas collisions. In the self-consistent
treatment of the energy balance in the circumstellar envelope a value
of 0.03 was assumed for $\langle Q\rangle$, which is also retained
here.  Photodissocation rates are taken from \citet{vanDishoeck1988},
with the assumption that SiS has the same photodissociation rate as CS
(confirmed to within a factor 2 by the UMIST Ratefile{\footnote{\tt
http://www.rate99.co.uk}}). The calculated shielding distances and
rate coefficients are shown in Tables \ref{tabmlr} and
\ref{tabphdist}, respectively.  

\begin{landscape}
\begin{table}
\begin{flushleft}
\caption{Abundances and sizes of emission regions$^a$}
\label{tababsiz}
\footnotesize
\begin{tabular}{l rrr r rrr r rrr r rrr}
\hline\hline
Molecule  &\multicolumn{3}{c}{IRC+10216 (SEST)}&&\multicolumn{3}{c}{IRC+10216 (OSO)}&&\multicolumn{3}{c}{IRAS15194-5115}&& \multicolumn{3}{c}{IRAS15082-4808}\\
\cline{2-4}\cline{6-8}\cline{10-12}\cline{14-16}
          &\multicolumn{1}{c}{$r_{\rm i}$}&\multicolumn{1}{c}{$r_{\rm e}$}&\multicolumn{1}{c}{$f_{\rm X}$}&
&\multicolumn{1}{c}{$r_{\rm i}$}&\multicolumn{1}{c}{$r_{\rm e}$}&\multicolumn{1}{c}{$f_{\rm X}$}&
&\multicolumn{1}{c}{$r_{\rm i}$}&\multicolumn{1}{c}{$r_{\rm e}$}&\multicolumn{1}{c}{$f_{\rm X}$}&
&\multicolumn{1}{c}{$r_{\rm i}$}&\multicolumn{1}{c}{$r_{\rm e}$}&\multicolumn{1}{c}{$f_{\rm X}$} \\
          &\multicolumn{1}{c}{[cm]}&\multicolumn{1}{c}{[cm]}&\multicolumn{1}{c}{[X]/[H$_2$]}&
&\multicolumn{1}{c}{[cm]}&\multicolumn{1}{c}{[cm]}&\multicolumn{1}{c}{[X]/[H$_2$]}&
&\multicolumn{1}{c}{[cm]}&\multicolumn{1}{c}{[cm]}&\multicolumn{1}{c}{[X]/[H$_2$]}&
&\multicolumn{1}{c}{[cm]}&\multicolumn{1}{c}{[cm]}&\multicolumn{1}{c}{[X]/[H$_2$]} \\
\hline
CN(1--0)                 & 2.6\,(16)	&7.1\,(16)	&{\bf 3.4\,(-06)}       &&	2.6\,(16)	&	7.1\,(16)	&	2.1\,(-06)         &&	3.5\,(16)	&	9.6\,(16)	&	2.1\,(-06)	        &&	2.0\,(16)	&	6.2\,(16)	&	3.2\,(-06)\\
$^{13}$CN(1--0)          & 2.6\,(16)	&7.1\,(16)	&7.6\,(-08)                &&	2.6\,(16)	&	7.1\,(16)	&	4.6\,(-08)         &&	3.5\,(16)	&	9.6\,(16)	&	3.0\,(-07)	        &&	2.0\,(16)	&	6.2\,(16)	&	$<$\,1.3\,(-08)	\\
CN(2--1)                 & 2.6\,(16)	&7.1\,(16)	&9.6\,(-07)                &&	2.6\,(16)	&	7.1\,(16)	&	---             &&	3.5\,(16)	&	9.6\,(16)	&	3.5\,(-07)	        &&	2.0\,(16)	&	6.2\,(16)	&	3.8\,(-07)\\
CS(2--1)                 &	        &4.0\,(16)	&{\bf 9.9\,(-07)}        &&		        &	4.0\,(16)	&	4.6\,(-07)         &&		        &	5.5\,(16)	&	2.3\,(-06)	        &&		        &	3.3\,(16)	&	2.2\,(-06)\\
$^{13}$CS(2--1)          &	        &4.0\,(16)	&2.2\,(-08)                &&		        &	4.0\,(16)	&	---             &&		        &	5.5\,(16)	&	5.0\,(-07)	        &&		        &	3.3\,(16)	&	$<$\,3.1\,(-08)	\\
C$^{34}$S(2--1)          &	        &4.0\,(16)	&3.7\,(-08)                &&		        &     	4.0\,(16)	&	---             &&		        &	5.5\,(16)	&	---             &&		        &	3.3\,(16)	&	--- \\
CS(5--4)                 &	        &4.0\,(16)	&1.2\,(-06)                &&		        &	4.0\,(16)	&	---             &&		        &	5.5\,(16)	&	1.6\,(-06)	        &&		        &	3.3\,(16)	&	9.0\,(-07)\\
C$_2$S(6,7--5,6)         & 1.2\,(16)    &4.0\,(16)      &$<$\,4.6\,(-09)            && 1.2\,(16)         &       4.0\,(16)       &  $<$\,4.5\,(-09)       && 8.0\,(15) & 5.5\,(16) & $<$\,2.4\,(-08) && 7.4\,(15) & 3.3\,(16) & $<$\,3.1\,(-08) \\
C$_2$S(8,9--7,8)         & 1.2\,(16)    &4.0\,(16)      &$<$\,4.1\,(-09)            && 1.2\,(16)         &       4.0\,(16)       &  $<$\,9.5\,(-09)       && 8.0\,(15) & 5.5\,(16) & $<$\,3.5\,(-08) && 7.4\,(15) & 3.3\,(16) & $<$\,4.8\,(-08) \\
C$_3$S(15-14)            & 1.2\,(16)    &4.0\,(16)      &$<$\,1.2\,(-08)            && 1.2\,(16)         &       4.0\,(16)       &  $<$\,1.5\,(-08)       && 8.0\,(15) & 5.5\,(16) & $<$\,7.2\,(-08) && 7.4\,(15) & 3.3\,(16) & $<$\,1.3\,(-06) \\
SiO(2--1)                &	        &2.0\,(16)	&1.3\,(-07)                &&		        &	2.0\,(16)	&	9.6\,(-08)         &&		        &	1.3\,(16)	&	1.7\,(-06)	        &&		        &	3.3\,(16)	&	7.2\,(-07)\\
SiS(5--4)                &	        &2.0\,(16)	&1.2\,(-06)                &&		        &	2.0\,(16)	&	8.1\,(-07)         &&		        &	1.3\,(16)	&	4.9\,(-06)	        &&		        &	3.3\,(16)	&	3.3\,(-06)\\
SiS(6--5)                &	        &2.0\,(16)	&9.0\,(-07)                &&		        &	2.0\,(16)	&	9.0\,(-07)         &&		        &	1.3\,(16)	&	4.2\,(-06)	        &&		        &	3.3\,(16)	&	2.0\,(-06)\\
SO(3--2)                & 2.6\,(16)	&7.1\,(16)      &$<$\,3.5\,(-09)            &&	2.6\,(16)	&	7.1\,(16)	&	---             &&	3.5\,(16)       &	9.6\,(16)      &	---                     &&	2.0\,(16)       &	6.2\,(16)       &	--- \\
HCN(1--0)         &	        &3.4\,(16)      &{\bf 1.4\,(-05)}                &&		        &	3.4\,(16)	&	{\bf 1.1\,(-05)}         &&		        &	4.6\,(16)	&	{\bf 1.2\,(-05)}	        &&		        &	2.7\,(16)	&	{\bf 1.0\,(-05)}\\
H$^{13}$CN(1--0)         &	        &3.4\,(16)      &3.1\,(-07)                &&		        &	3.4\,(16)	&	2.5\,(-07)         &&		        &	4.6\,(16)	&	2.0\,(-06)	        &&		        &	2.7\,(16)	&	2.9\,(-07)\\
HNC(1--0)                & 2.4\,(16)	&8.4\,(16)	&7.2\,(-08)                &&	2.4\,(16)	&	8.4\,(16)	&	3.8\,(-08)         &&	1.6\,(16)	&	5.6\,(16)	&	3.3\,(-07)	        &&	1.5\,(16)	&	5.2\,(16)	&	1.6\,(-07)\\
HN$^{13}$C(1--0)         & 2.4\,(16)	&8.4\,(16)	&$<$\,1.9\,(-09)            &&	2.4\,(16)	&	8.4\,(16)	&	---             &&	1.6\,(16)	&	5.6\,(16)	&	1.1\,(-07)	        &&	1.5\,(16)	&	5.2\,(16)	&	$<$\,1.2\,(-08)	\\
SiC$_2$(4,04--3,03)      & 2.4\,(16)	&6.0\,(16)	&1.6\,(-07)                &&	2.4\,(16)	&	6.0\,(16)	&	---             &&	1.6\,(16)	&	4.0\,(16)	&	---             &&	1.5\,(16)	&	3.7\,(16)	&	3.3\,(-07)\\
SiC$_2$(4,22--3,21)      & 2.4\,(16)	&6.0\,(16)	&3.2\,(-07)                &&	2.4\,(16)	&	6.0\,(16)	&	---             &&	1.6\,(16)	&	4.0\,(16)	&	---             &&	1.5\,(16)	&	3.7\,(16)	&	$<$\,1.8\,(-07)	\\
SiC$_2$(5,05--4,04)      & 2.4\,(16)	&6.0\,(16)	&1.7\,(-07)                &&	2.4\,(16)	&	6.0\,(16)	&	1.7\,(-07)         &&	1.6\,(16)	&	4.0\,(16)	&	1.2\,(-06)	        &&	1.5\,(16)	&	3.7\,(16)	&	4.9\,(-07)\\
C$_2$H(1--0)             & 2.3\,(16)	&5.6\,(16)	&2.8\,(-06)                &&	2.3\,(16)	&	5.6\,(16)	&	2.4\,(-06)         &&	1.6\,(16)	&	4.8\,(16)	&	1.5\,(-05)	        &&	1.7\,(16)	&	4.8\,(16)	&	8.3\,(-06)\\
C$_3$H(9/2--7/2)         & 2.4\,(16)	&8.4\,(16)	&5.5\,(-08)                &&	2.4\,(16)	&	8.4\,(16)	&	5.9\,(-08)         &&	1.6\,(16)	&	5.6\,(16)	&	9.2\,(-08)	        &&	1.5\,(16)	&	5.2\,(16)	&	$<$\,1.7\,(-08)	\\
C$_3$N(9--8)             & 2.4\,(16)	&8.4\,(16)	&3.0\,(-07)                &&	2.4\,(16)	&	8.4\,(16)	&	---             &&	1.6\,(16)	&	5.6\,(16)	&	$<$\,4.9\,(-08)	&&	1.5\,(16)	&	5.2\,(16)	&	$<$\,1.0\,(-07)	\\
C$_3$N(11--10)           & 2.4\,(16)	&8.4\,(16)	&5.9\,(-07)                &&	2.4\,(16)	&	8.4\,(16)	&	7.6\,(-07)         &&	1.6\,(16)	&	5.6\,(16)	&	9.2\,(-07)	        &&	1.5\,(16)	&	5.2\,(16)	&	9.5\,(-07)\\
C$_4$H(9--8)             & 2.4\,(16)	&8.4\,(16)	&3.7\,(-06)                &&	2.4\,(16)	&	8.4\,(16)	&	---             &&	1.6\,(16)	&	5.6\,(16)	&	---             &&	1.5\,(16)	&	5.2\,(16)	&	$<$\,6.6\,(-07)	\\
C$_4$H(10--9)            & 2.4\,(16)	&8.4\,(16)	&2.7\,(-06)                &&	2.4\,(16)	&	8.4\,(16)	&	3.2\,(-06)         &&	1.6\,(16)	&	5.6\,(16)	&	2.8\,(-05)	        &&	1.5\,(16)	&	5.2\,(16)	&	6.9\,(-06)\\
C$_3$H$_2$(2,12--1,01)   & 2.4\,(16)	&8.4\,(16)	&3.2\,(-08)                &&	2.4\,(16)	&	8.4\,(16)	&	---             &&	1.6\,(16)	&	5.6\,(16)	&	6.0\,(-07)	        &&	1.5\,(16)	&	5.2\,(16)	&	$<$\,9.3\,(-08)	\\
HC$_3$N(10--9)           & 1.2\,(16)	&6.0\,(16)	&{\bf 1.1\,(-06)}       &&	1.2\,(16)	&	6.0\,(16)	&	4.4\,(-07)         &&	8.0\,(15)	&	4.0\,(16)	&	1.9\,(-06)	        &&	7.4\,(15)	&	3.7\,(16)	&	2.3\,(-06)\\
HCC$^{13}$CN(10--9)      & 1.2\,(16)	&6.0\,(16)	&2.4\,(-08)                &&	1.2\,(16)	&	6.0\,(16)	&	$<$\,2.8\,(-09)	&&	8.0\,(15)	&	4.0\,(16)	&	3.9\,(-07)	        &&	7.4\,(15)	&	3.7\,(16)	&	$<$\,2.7\,(-08)	\\
HC$^{13}$CCN(10--9)      & 1.2\,(16)	&6.0\,(16)	&2.4\,(-08)                &&	1.2\,(16)	&	6.0\,(16)	&	$<$\,2.8\,(-09)	&&	8.0\,(15)	&	4.0\,(16)	&	3.9\,(-07)	        &&	7.4\,(15)	&	3.7\,(16)	&	$<$\,2.7\,(-08)	\\
HC$_3$N(11--10)          & 1.2\,(16)	&6.0\,(16)	&7.5\,(-07)                &&	1.2\,(16)	&	6.0\,(16)	&	---             &&	8.0\,(15)	&	4.0\,(16)	&	---             &&	7.4\,(15)	&	3.7\,(16)	&	--- \\
HC$_3$N(12--11)          & 1.2\,(16)	&6.0\,(16)	&{\bf 1.9\,(-06)}       &&	1.2\,(16)	&	6.0\,(16)	&	---             &&	8.0\,(15)	&	4.0\,(16)	&	2.2\,(-06)	        &&	7.4\,(15)	&	3.7\,(16)	&	2.1\,(-06)\\
HCC$^{13}$CN(12--11)     & 1.2\,(16)	&6.0\,(16)	&4.1\,(-08)                &&	1.2\,(16)	&	6.0\,(16)	&	1.5\,(-08)         &&	8.0\,(15)	&	4.0\,(16)	&	2.8\,(-07)	        &&	7.4\,(15)	&	3.7\,(16)	&	$<$\,2.0\,(-08)	\\
HC$^{13}$CCN(12--11)     & 1.2\,(16)	&6.0\,(16)	&4.1\,(-08)                &&	1.2\,(16)	&	6.0\,(16)	&	1.5\,(-08)         &&	8.0\,(15)	&	4.0\,(16)	&	2.8\,(-07)	        &&	7.4\,(15)	&	3.7\,(16)	&	$<$\,2.0\,(-08)	\\
CH$_3$CN(6(1)--5(1))     & 1.2\,(16)	&6.0\,(16)	&1.2\,(-08)                &&	1.2\,(16)	&	6.0\,(16)	&	$<$\,2.0\,(-09)	&&	8.0\,(15)	&	4.0\,(16)	&	$<$\,2.5\,(-08)	&&	7.4\,(15)	&	3.7\,(16)	&	$<$\,4.3\,(-08)	\\
CH$_3$CN(12(0)--11(0))     & 1.2\,(16)	&6.0\,(16)	&$<$\,1.2\,(-07)                &&	1.2\,(16)	&	6.0\,(16)	& ---	&&	8.0\,(15)	&	4.0\,(16)	&	$<$\,2.4\,(-06)	&&	7.4\,(15)	&	3.7\,(16)	&	$<$\,1.5\,(-06)	\\
HC$_5$N(32--31)          & 1.2\,(16)	&6.0\,(16)	&3.9\,(-06)                &&	1.2\,(16)	&	6.0\,(16)	&	---             &&	8.0\,(15)	&	4.0\,(16)	&	$<$\,4.5\,(-06)	&&	7.4\,(15)	&	3.7\,(16)	&	$<$\,3.8\,(-06)	\\
HC$_5$N(34--33)          & 1.2\,(16)	&6.0\,(16)	&4.2\,(-06)                &&	1.2\,(16)	&	6.0\,(16)	&	$<$\,7.6\,(-07)	&&	8.0\,(15)	&	4.0\,(16)	&	---             &&	7.4\,(15)	&	3.7\,(16)	&	$<$\,7.4\,(-06)	\\
HC$_5$N(35--34)          & 1.2\,(16)	&6.0\,(16)	&1.3\,(-05)                &&	1.2\,(16)	&	6.0\,(16)	&	---             &&	8.0\,(15)	&	4.0\,(16)	&	---             &&	7.4\,(15)	&	3.7\,(16)	&	$<$\,1.1\,(-05)	\\
\hline
\end{tabular}
  {\footnotesize
   \begin{enumerate}
 	\renewcommand{\labelenumi}{(\alph{enumi})}
	\item In this table $x\,(y)$\ represents $x\times10^y$.
	\item Bold face indicates an abundance calculated by scaling
	the abundance of the $^{13}$C isotope of the same species by
	the modelled $^{12}$CO/$^{13}$CO-ratio (see
	Sect.~\ref{distance}).
   \end{enumerate}
     }
\normalsize
\end{flushleft}
\end{table}
\end{landscape}
\begin{landscape}
\begin{table}
\begin{flushleft}
\addtocounter{table}{-1}
\caption{(\emph{cont.}) Abundances and sizes of emission regions$^a$}
\footnotesize
\begin{tabular}{l rrr r rrr r rrr r rrr}
\hline\hline
Molecule  &\multicolumn{3}{c}{IRAS07454-7112}&&\multicolumn{3}{c}{CIT\,6}&&\multicolumn{3}{c}{AFGL\,3068} &&\multicolumn{3}{c}{IRC+40540}\\
\cline{2-4}\cline{6-8}\cline{10-12}\cline{14-16}
          &\multicolumn{1}{c}{$r_{\rm i}$}&\multicolumn{1}{c}{$r_{\rm e}$}&\multicolumn{1}{c}{$f_{\rm X}$}&
&\multicolumn{1}{c}{$r_{\rm i}$}&\multicolumn{1}{c}{$r_{\rm e}$}&\multicolumn{1}{c}{$f_{\rm X}$}&
&\multicolumn{1}{c}{$r_{\rm i}$}&\multicolumn{1}{c}{$r_{\rm e}$}&\multicolumn{1}{c}{$f_{\rm X}$}&
&\multicolumn{1}{c}{$r_{\rm i}$}&\multicolumn{1}{c}{$r_{\rm e}$}&\multicolumn{1}{c}{$f_{\rm X}$} \\
          &\multicolumn{1}{c}{cm}&\multicolumn{1}{c}{cm}&\multicolumn{1}{c}{[X]/[H$_2$]}&
&\multicolumn{1}{c}{cm}&\multicolumn{1}{c}{cm}&\multicolumn{1}{c}{[X]/[H$_2$]}&
&\multicolumn{1}{c}{cm}&\multicolumn{1}{c}{cm}&\multicolumn{1}{c}{[X]/[H$_2$]}&
&\multicolumn{1}{c}{cm}&\multicolumn{1}{c}{cm}&\multicolumn{1}{c}{[X]/[H$_2$]} \\
\hline
CN(1--0)                 & 1.3\,(16)	&	4.1\,(16)	&	5.3\,(-06)         &&	1.6\,(16)	&	5.1\,(16)	&	{\bf 2.0\,(-05)}        &&	3.2\,(16)	&	8.3\,(16)	&	4.6\,(-07)         &&	3.2\,(16)	&	8.3\,(16)	&       1.0\,(-06)	\\	
$^{13}$CN(1--0)          & 1.3\,(16)	&	4.1\,(16)	&	4.8\,(-07)         &&	1.6\,(16)	&	5.1\,(16)	&	5.8\,(-07)                 &&	3.2\,(16)	&	8.3\,(16)	&	$<$\,2.6\,(-08)	&&	3.2\,(16)	&	8.3\,(16)	&	$<$\,9.9\,(-09)	\\
CN(2--1)                 & 1.3\,(16)	&	4.1\,(16)	&	1.6\,(-06)         &&	1.6\,(16)	&	5.1\,(16)	&	---                     &&	3.2\,(16)	&	8.3\,(16)	&	---             &&	3.2\,(16)	&	8.3\,(16)	&	---	\\
CS(2--1)                 &	        &	2.2\,(16)	&	1.6\,(-06)         &&		        &	2.7\,(16)	&	2.5\,(-06)                 &&		        &	4.8\,(16)	&	3.7\,(-07)         &&		        &	4.9\,(16)	&	5.4\,(-07)	\\
$^{13}$CS(2--1)          &	        &	2.2\,(16)	&	$<$\,2.9\,(-08)	&&		        &	2.7\,(16)	&	---                     &&		        &	4.8\,(16)	&	---             &&		        &	4.9\,(16)	&	---	\\
C$^{34}$S (2--1)         &	        &	2.2\,(16)	&	---             &&		        &	2.7\,(16)	&	---                     &&		        &	4.8\,(16)	&	---             &&		        &	4.9\,(16)	&	---	\\
CS(5--4)                 &	        &	2.2\,(16)	&	1.6\,(-06)         &&		        &	2.7\,(16)	&	---                     &&		        &	4.8\,(16)	&	---             &&		        &	4.9\,(16)	&	---	\\
C$_2$S(6,7--5,6)         & 5.6\,(15)    &2.2\,(16)      &$<$\,9.1\,(-08)            && 4.3\,(15)         &       2.7\,(16)       &  $<$\,1.0\,(-07)       && 1.6\,(16) & 4.8\,(16) & $<$\,8.1\,(-08) && 1.6\,(16) & 4.9\,(16) & $<$\,3.0\,(-08) \\
C$_2$S(8,9--7,8)         & 5.6\,(15)    &2.2\,(16)      &$<$\,1.1\,(-07)            && 4.3\,(15)         &       2.7\,(16)       &  $<$\,2.0\,(-07)       && 1.6\,(16) & 4.8\,(16) & $<$\,8.0\,(-08) && 1.6\,(16) & 4.9\,(16) & $<$\,5.6\,(-08) \\
C$_3$S(15-14)            & 5.6\,(15)    &2.2\,(16)      &$<$\,3.6\,(-07)            && 4.3\,(15)         &       2.7\,(16)       &  $<$\,2.5\,(-07)       && 1.6\,(16) & 4.8\,(16) & $<$\,3.5\,(-07) && 1.6\,(16) & 4.9\,(16) & $<$\,8.9\,(-08) \\
SiO(2--1)                &	        &	9.4\,(15)	&	4.4\,(-07)         &&		        &	7.2\,(15)	&	1.0\,(-06)                 &&		        &	2.6\,(16)	&	$<$\,4.7\,(-08)	&&		        &	2.6\,(16)	&	5.6\,(-08)	\\
SiS(5--4)                &	        &	9.4\,(15)	&	4.8\,(-06)         &&		        &	7.2\,(15)	&	1.0\,(-06)                 &&		        &	2.6\,(16)	&	3.3\,(-07)         &&		        &	2.6\,(16)	&	1.4\,(-06)	\\
SiS(6--5)                &	        &	9.4\,(15)	&	3.4\,(-06)         &&		        &	7.2\,(15)	&	5.2\,(-06)                 &&		        &	2.6\,(16)	&	1.0\,(-06)         &&		        &	2.6\,(16)	&	4.9\,(-07)	\\
HCN(1--0)         &	        &	1.8\,(16)	&	{\bf 7.8\,(-06)}           &&		        &	2.2\,(16)	&	{\bf 1.1\,(-05)}          &&		        &	4.2\,(16)	&	{\bf 6.3\,(-06)}         &&		        &	4.2\,(16)	&	{\bf 6.5\,(-06)}	\\
H$^{13}$CN(1--0)         &	        &	1.8\,(16)	&	5.6\,(-07)         &&		        &	2.2\,(16)	&	3.0\,(-07)                 &&		        &	4.2\,(16)	&	2.1\,(-07)         &&		        &	4.2\,(16)	&	1.3\,(-07)	\\
HNC(1--0)                & 1.1\,(16)	&	4.0\,(16)	&	1.0\,(-07)         &&	8.6\,(15)	&	3.0\,(16)	&	2.3\,(-07)                 &&	3.1\,(16)	&	1.1\,(17)	&	3.0\,(-08)         &&	3.1\,(16)	&	1.1\,(17)	&	2.2\,(-08)	\\
HN$^{13}$C(1--0)         & 1.1\,(16)	&	4.0\,(16)	&	$<$\,1.8\,(-08)	&&	8.6\,(15)	&	3.0\,(16)	&	$<$\,2.1\,(-08)	                &&	3.1\,(16)	&	1.1\,(17)	&	$<$\,1.4\,(-08)	&&	3.1\,(16)	&	1.1\,(17)	&	$<$\,4.4\,(-09)	\\
SiC$_2$(4,04--3,03)      & 1.1\,(16)	&	2.8\,(16)	&	2.3\,(-07)         &&	8.6\,(15)	&	2.2\,(16)	&	---                     &&	3.1\,(16)	&	7.8\,(16)	&	---             &&	3.1\,(16)	&	7.8\,(16)	&	---	\\
SiC$_2$(4,22--3,21)      & 1.1\,(16)	&	2.8\,(16)	&	$<$\,2.7\,(-07)	        &&	8.6\,(15)	&	2.2\,(16)	&	---                     &&	3.1\,(16)	&	7.8\,(16)	&	---             &&	3.1\,(16)	&	7.8\,(16)	&	---	\\
SiC$_2$(5,05--4,04)      & 1.1\,(16)	&	2.8\,(16)	&	$<$\,3.2\,(-07)	&&	8.6\,(15)	&	2.2\,(16)	&	3.1\,(-06)                 &&	3.1\,(16)	&	7.8\,(16)	&	$<$\,7.5\,(-08)	&&	3.1\,(16)	&	7.8\,(16)	&	$<$\,5.3\,(-08)	\\
C$_2$H(1--0)             & 1.1\,(16)	&	3.2\,(16)	&	$<$\,4.3\,(-07)	&&	1.4\,(16)	&	4.1\,(16)	&	6.9\,(-06)                 &&	2.8\,(16)	&	6.7\,(16)	&	5.7\,(-06)         &&	2.7\,(16)	&	6.7\,(16)	&	$<$\,1.0\,(-07)	\\
C$_3$H(9/2--7/2)         & 1.1\,(16)	&	4.0\,(16)	&	$<$\,3.1\,(-08)	&&	8.6\,(15)	&	3.0\,(16)	&	$<$\,7.4\,(-08)	        &&	3.1\,(16)	&	1.1\,(17)	&	$<$\,1.4\,(-08)	&&	3.1\,(16)	&	1.1\,(17)	&	$<$\,1.2\,(-08)	\\
C$_3$N(9--8)             & 1.1\,(16)	&	4.0\,(16)	&	$<$\,1.6\,(-07)	&&	8.6\,(15)	&	3.0\,(16)	&	---                     &&	3.1\,(16)	&	1.1\,(17)	&	---             &&	3.1\,(16)	&	1.1\,(17)	&	---	\\
C$_3$N(11--10)           & 1.1\,(16)	&	4.0\,(16)	&	7.3\,(-07)         &&	8.6\,(15)	&	3.0\,(16)	&	2.6\,(-06)                 &&	3.1\,(16)	&	1.1\,(17)	&	5.5\,(-07)         &&	3.1\,(16)	&	1.1\,(17)	&	1.1\,(-07)	\\
C$_4$H(9--8)             & 1.1\,(16)	&	4.0\,(16)	&	---             &&	8.6\,(15)	&	3.0\,(16)	&	---                     &&	3.1\,(16)	&	1.1\,(17)	&	---             &&	3.1\,(16)	&	1.1\,(17)	&	---	\\
C$_4$H(10--9)            & 1.1\,(16)	&	4.0\,(16)	&	$<$\,8.8\,(-07)	&&	8.6\,(15)	&	3.0\,(16)	&	$<$\,1.7\,(-06)	        &&	3.1\,(16)	&	1.1\,(17)	&	---             &&	3.1\,(16)	&	1.1\,(17)	&	---	\\
C$_3$H$_2$(2,12--1,01)   & 1.1\,(16)	&	4.0\,(16)	&	---             &&	8.6\,(15)	&	3.0\,(16)	&	---                     &&	3.1\,(16)	&	1.1\,(17)	&	---             &&	3.1\,(16)	&	1.1\,(17)	&	---	\\
HC$_3$N(10--9)           & 5.6\,(15)	&	2.8\,(16)	&	1.7\,(-06)         &&	4.3\,(15)	&	2.2\,(16)	&	2.4\,(-06)                 &&	1.6\,(16)	&	7.8\,(16)	&	5.0\,(-07)         &&	1.6\,(16)	&	7.8\,(16)	&	{\bf 1.5\,(-06)}	\\
HCC$^{13}$CN(10--9)      & 5.6\,(15)	&	2.8\,(16)	&	1.5\,(-07)         &&	4.3\,(15)	&	2.2\,(16)	&	$<$\,4.9\,(-08)	        &&	1.6\,(16)	&	7.8\,(16)	&	$<$\,1.5\,(-08)	&&	1.6\,(16)	&	7.8\,(16)	&	3.1\,(-08)	\\
HC$^{13}$CCN(10--9)      & 5.6\,(15)	&	2.8\,(16)	&	1.5\,(-07)         &&	4.3\,(15)	&	2.2\,(16)	&	$<$\,4.9\,(-08)	        &&	1.6\,(16)	&	7.8\,(16)	&	$<$\,1.5\,(-08)	&&	1.6\,(16)	&	7.8\,(16)	&	3.1\,(-08)	\\
HC$_3$N(11--10)          & 5.6\,(15)	&	2.8\,(16)	&	---             &&	4.3\,(15)	&	2.2\,(16)	&	---                     &&	1.6\,(16)	&	7.8\,(16)	&	---             &&	1.6\,(16)	&	7.8\,(16)	&	---	\\
HC$_3$N(12--11)          & 5.6\,(15)	&	2.8\,(16)	&	2.4\,(-06)         &&	4.3\,(15)	&	2.2\,(16)	&	---                     &&	1.6\,(16)	&	7.8\,(16)	&	---             &&	1.6\,(16)	&	7.8\,(16)	&	---	\\
HCC$^{13}$CN(12--11)     & 5.6\,(15)	&	2.8\,(16)	&	1.3\,(-07)         &&	4.3\,(15)	&	2.2\,(16)	&	2.7\,(-07)                 &&	1.6\,(16)	&	7.8\,(16)	&	$<$\,2.9\,(-08)	&&	1.6\,(16)	&	7.8\,(16)	&	$<$\,1.1\,(-08)	\\
HC$^{13}$CCN(12--11)     & 5.6\,(15)	&	2.8\,(16)	&	1.3\,(-07)         &&	4.3\,(15)	&	2.2\,(16)	&	2.7\,(-07)                 &&	1.6\,(16)	&	7.8\,(16)	&	$<$\,2.9\,(-08)	&&	1.6\,(16)	&	7.8\,(16)	&	$<$\,1.1\,(-08)	\\
CH$_3$CN(6(1)--5(1))           & 5.6\,(15)	&	2.8\,(16)	&	$<$\,2.4\,(-07)	&&	4.3\,(15)	&	2.2\,(16)	&	$<$\,1.3\,(-07)	        &&	1.6\,(16)	&	7.8\,(16)	&	$<$\,1.2\,(-07)	&&	1.6\,(16)	&	7.8\,(16)	&	$<$\,6.4\,(-08)	\\
CH$_3$CN(12(0)--11(0))           & 5.6\,(15)	&	2.8\,(16)	&	$<$\,4.6\,(-06)	&&	4.3\,(15)	&	2.2\,(16)	&	---	        &&	1.6\,(16)	&	7.8\,(16)	&	---	&&	1.6\,(16)	&	7.8\,(16)	&	---	\\
HC$_5$N(32--31)          & 5.6\,(15)	&	2.8\,(16)	&	---             &&	4.3\,(15)	&	2.2\,(16)	&	---                     &&	1.6\,(16)	&	7.8\,(16)	&	---             &&	1.6\,(16)	&	7.8\,(16)	&	---	\\
HC$_5$N(34--33)          & 5.6\,(15)	&	2.8\,(16)	&	9.2\,(-06)         &&	4.3\,(15)	&	2.2\,(16)	&	$<$\,1.3\,(-05)	        &&	1.6\,(16)	&	7.8\,(16)	&	$<$\,4.1\,(-06)	&&	1.6\,(16)	&	7.8\,(16)	&	$<$\,4.5\,(-06)	\\
HC$_5$N(35--34)          & 5.6\,(15)	&	2.8\,(16)	&	$<$\,8.8\,(-06)	&&	4.3\,(15)	&	2.2\,(16)	&	---                     &&	1.6\,(16)	&	7.8\,(16)	&	---             &&	1.6\,(16)	&	7.8\,(16)	&	---	\\
\hline
\end{tabular}
\end{flushleft}
  {\footnotesize
   \begin{enumerate}
 	\renewcommand{\labelenumi}{(\alph{enumi})}
	\item In this table $x\,(y)$\ represents $x\times10^y$.
	\item Bold face indicates an abundance calculated by scaling
	the abundance of the $^{13}$C isotope of the same species by
	the modelled $^{12}$CO/$^{13}$CO-ratio (see
	Sect.~\ref{distance}).
   \end{enumerate}
     }
\normalsize
\end{table}
\end{landscape}

The photodissociation radii of species formed in the photosphere were
chosen to be the e--folding radii of the initial abundances,
i.e. $f_{\mathrm{X}}(r_{\mathrm{e}}) =
f_{\mathrm{X}}(R_{\star})/{\mathrm{e}}$, where $R_{\star}$\ is the
stellar radius. For the photodissociation products inner and outer
radii are chosen to be where the abundance has dropped to 1/e of its
peak value.  The envelope sizes calculated by the simple
photodissociation model agree relatively well with observed values in
IRC+10216 and other objects, except in the case of SiS (see next
paragraph).  \citet{Lindqvist_etal2000} used a detailed radiative
transfer method and observed brightness distributions to derive
envelope sizes of HCN and CN. For \object{IRC+10216} and
\object{IRC+40540} they found envelope sizes of
$\sim$4$\times$10$^{16}$\,cm for HCN and $\sim$5$\times$10$^{16}$\,cm
for CN, in excellent agreement with the values derived here from the
photodissociation model. In the case of \object{CIT 6} the observed
HCN envelope size is larger than that calculated by a factor of
2. Envelope sizes and derived abundances are given in Table
\ref{tababsiz}.

SiO and SiS are parent species with a radial extent depending on
photodissociation. The calculation of the SiS photodissociation radius
in \object{IRC+10216}, however, is not consistent with the radius
observed by \citet{BiegingTafalla1993}. Therefore the observed radius
for SiS in \object{IRC+10216} was used in the abundance calculations,
and its radius was scaled for the other objects in the sample with the
factor
$(\dot{M}/v_{\mathrm{exp}})_{\star}/(\dot{M}/v_{\mathrm{exp}})_{10216}$\ in
the same way as described in section~\ref{chemrad}. The outer radius
calculated via the photodissociation model is actually slightly larger
than that observed. This gives credence to the idea that SiS freezes
out onto dust grains. SiO is treated in the same way as SiS since it
too is likely to freeze out onto grains. Hence the same envelope size
as SiS was adopted for this species.

\subsubsection{Inner and outer radii for species of circumstellar chemistry 
origin}
\label{chemrad}
For those molecules that have their origin in a circumstellar
chemistry a different approach is used. The sizes of the emission
regions of HNC and HC$_3$N have been determined interferometrically
for \object{IRC+10216} using observations by \citet{BiegingRieu1988},
and of SiC$_2$ by \citet{Lucas_etal1995}.  Inner and outer radii for
HNC, HC$_3$N and SiC$_2$ are taken from observations of
\object{IRC+10216}, and scaled by a factor
$(\dot{M}/v_{\mathrm{exp}})_{\star}/(\dot{M}/v_{\mathrm{exp}})_{10216}$\ for
the remaining six sample carbon stars (since molecular distributions
have not been mapped in all the sample stars; see Table \ref{tabmlr}
for values of this ratio).  This factor is proportional to the density
ratio at a given radius, and also (to the first order) to the ratio of
the envelope shielding distances.  Chemical modelling shows that the
species C$_3$H, C$_4$H, C$_3$H$_2$, C$_3$N, CH$_3$CN, HC$_5$N, SO,
C$_2$S and C$_3$S are formed via chemical reactions in the outer
envelope \citep[e.g.,][]{MillarHerbst1994}.  Furthermore, it shows
that the species C$_3$H, C$_4$H, C$_3$H$_2$ and C$_3$N have a similar
radial distribution to HNC, and that the species CH$_3$CN and HC$_5$N
are similarly distributed to HC$_3$N. Hence corresponding radii for
these species are assumed in the calculations.  SO is assumed to have
a similar distribution to CN \citep[c.f.,][]{NejadMillar1988}, and
C$_2$S and C$_3$S are assumed to follow the CS distribution
\citep[c.f.,][]{Millar_etal2001}, although these latter two species
are not parent molecules (and hence are given an inner radius equal to
that of HC$_3$N).  All isotopes (i.e. species involving $^{13}$C and
$^{34}$S) are further assumed to have the same distributions as the
main isotope.

\begin{table*}
\begin{flushleft}
\caption{Molecular isotope abundance ratios}
\label{12c/13c}
\begin{tabular}{lcccccc}
\hline\hline
                             & IRC+10216 (S) & IRC+10216 (O) & IRAS15194 & IRAS07454 & CIT\,6   & IRC+40540 \\
\hline
HC$_3$N/HC$^{13}$CCN(10--9)  &	32.6 	        &	---	  &	4.9   &	11.1      & ---      &	17.9	\\
HC$_3$N/HC$^{13}$CCN(12--11) &	22.1	        &	---	  &	7.9   &	17.8      & ---	     &	---	\\
CS/$^{13}$CS(2--1)	     &	22.7	        &	---	  &	4.6   &	---       & ---	     &	---	\\
CN/$^{13}$CN(1--0)	     &	25.1	        &	46.6	  &	6.7   &	11.0	  & 12.4     &	---	\\
\hline
Average	                     &	25.6	        &	46.6	  &	6.0   & 13.3      & 12.4     &	17.9	\\
\hline
Modelling of CO (Table~\ref{tabmlr}) &  45.0            &       45.0      &     6.0   & 14.0      & 35.0     &  50.0 \\
\hline
\end{tabular}
\end{flushleft}
\end{table*}

\subsection{Molecular abundances}
\label{distance}

\subsubsection{Uncertainties in the abundance estimates}

The assumption of optically thin emission has a systematic effect on
the derived abundances, in that the true abundance will be higher if
there are opacity effects.  As can be seen from Table~\ref{12c/13c},
there can be a factor of 2--3 in error from this assumption.

The accuracy of the calculated abundances is dependent on various
assumptions. Due to the intrinsic difficulties in estimating distances
to the objects included in the sample, the typical uncertainty in the
adopted distance is a factor of $\sim$2. This influences the mass-loss
rates derived in the radiative transfer modelling such that the
adopted mass-loss rate will scale as $D^{1-2}$. The calculation of the
limiting radii of a certain molecular distribution in the envelope is
also affected by inaccuracies in the distance estimate. Where radii
have been calculated by scaling observed radii in \object{IRC+10216},
there comes an error which scales as $D$. The results from the
photodissociation model will have a lesser dependence, with $D$\
coming into the expression for the shielding distance, via
$\dot{M}$. Hence the abundance, which is trivially derived from
Eq.~\ref{optthin}, will vary as approximately $D^{-1-0}$, giving
a factor of 2, possibly, in error.  

In the present analysis an excitation temperature of 10\,K was assumed
for all transitions. The error in abundance estimate due to this
assumption will depend on the excitation temperature of a particular 
transition and the assumption of LTE, and is estimated to be not more 
than a factor of $\sim$2.

Overall, it seems like an error of a factor of 5 is reasonable to
expect in the abundances presented here, although it should be borne
in mind that this could possibly rise to an order of magnitude. In the
comparison of abundance ratios in Table~\ref{tababrat}, any difference
less than a factor of 5 is treated as insignificant with respect to
error margins.

\subsubsection{Comparison with previously published abundances}

In comparison with \citet{Nyman_etal1993}, derived abundance ratios
(Table~\ref{tababrat}) have increased in favour of \object{IRAS
15194--5115} by up to a factor of approximately 4.  Individual
abundances generally show a factor $\sim$2 increase for those in
\object{IRAS 15914--5115}, and a factor $\sim$2 decrease for
\object{IRC+10216}, over \citet{Nyman_etal1993}.  In
\citet{Nyman_etal1993}, the distances adopted for \object{IRC+10216}
and \object{IRAS 15194-5115} were approximately twice as large as
those here; however this has little effect on calculated abundances
since the distance ratio is more or less unchanged.  However, the
calculated mass-loss rate of \object{IRAS 15194--5115} in the previous
paper was larger than that of \object{IRC+10216} by a factor of 2.5,
and here the recalculated mass-loss rates are the same for these two
objects.  This would tend to increase the abundance ratios quoted by a
similar factor. The photodissociation radius depends on the mass-loss
rate and the dust parameters through the dust shielding
distance. Compared to \citet{Nyman_etal1993} this paper uses different
mass-loss rates, a different gas-to-dust ratio (0.01 compared to
0.005), and the dust parameters are also slightly different due to the
determination of the $h$--factor in the radiative transfer
analysis. The scale factor that determines the inner and outer radii
of the species with a origin in circumstellar chemistry depends on the
ratio of the mass-loss rates. Thus the difference in relative mass
loss rates explains the factor of 4 difference in relative abundances
between \object{IRAS 15914--5115} and \object{IRC+10216} derived in
this paper compared to those derived in \citet{Nyman_etal1993}.

As discussed earlier in this paper the outer radii of SiS and SiO are
not determined through the photodissociation radius but scaled from
their observed radius in IRC+10216. In this way the calculated
abundances of SiO and SiS have increased in by a factor of 4--5 for
\object{IRAS 15194--5115} compared to \citet{Nyman_etal1993}.  For
\object{IRC+10216} these species have the same abundance as calculated
previously, and they are in reasonable agreement with those reported
elsewhere in the literature.  \citet{Bujarrabal_etal1994} give
$f($SiO$)$ = 5.6\,10$^{-7}$\ and $f($SiS$)$ = 3.9\,10$^{-6}$, which
are $\sim$4 times greater, using an outer radius half that quoted in
this paper and a distance of 200\,pc.  Better agreement is seen for
the other species which \citet{Bujarrabal_etal1994} detect, with the
exception of HNC, which is a factor $\sim$6 in disagreement.
\citet*{Cernicharo_etal2000} also calculate abundances in
\object{IRC+10216}, and are within a factor 5 of those here.

Generally, it seems that \object{IRC+10216} has lower
abundances than \object{IRAS 15194--5115}, \object{IRAS 15082--4808},
\object{IRAS 07454--7112} and \object{CIT 6}.  However, it is very
similar, physically and chemically, to \object{AFGL 3068} and
\object{IRC+40540}.

The three northern sources observed at OSO have also been studied in
the literature, and abundances derived. The
\citet{Bujarrabal_etal1994} paper includes data relating to
\object{CIT 6}, \object{AFGL 3068} and \object{IRC+40540}. All
abundances calculated by Bujarrabal et al. in these three sources are
greater than those derived here.  Distances and mass-loss rates are
reasonably comparable between this paper and that.  Generally, this
means that the calculated abundances in this paper are lower by a
factor of $\sim$2 in \object{CIT 6}, a factor of $\sim$5 in
\object{AFGL 3068}, and a factor of $\sim$3 in \object{IRC+40540}
compared to those derived in Bujarrabal et al. (1994).

\begin{table*}
\begin{flushleft}
\caption{Abundance ratios, compared to IRC+10216 observed with the SEST.}
\label{tababrat}
\begin{tabular}{l ccccccccc}
\hline\hline
                    & \multicolumn{3}{c}{IRC+10216} & IRAS15194 & IRAS15082 & IRAS07454 & \phantom{00}CIT\,6\phantom{00} & AFGL\,3068 & IRC+40540 \\
\cline{2-4}
Molecule                    & SEST && OSO \\
\hline
CN(1-0)                     & 1.0 && 1.1 & \phantom{0}1.1	&	1.7 &	2.8	& \phantom{0}3.8       & 0.2 & 0.5\\
$^{13}$CN(1-0)              & 1.0 && 0.6 & \phantom{0}4.0	&	--- & {\bf 6.3} & {\bf\phantom{0}7.6}  & --- & ---\\
CN(2-1)                     & 1.0 && --- & \phantom{0}0.4	&	0.4 &	1.7	&	---            & --- & ---\\
CS(2-1)                     & 1.0 && 0.9 & \phantom{0}4.6	&	4.3 &	3.2	& {\bf\phantom{0}5.1}  & 0.7 & 1.1\\
$^{13}$CS(2-1)              & 1.0 && --- & {\bf 22.7}           &	--- &	---	&	---            & --- & ---\\
CS(5-4)                     & 1.0 && --- & \phantom{0}1.4	&	0.8 &	1.4	&	---            & --- & ---\\
SiO(2-1)                    & 1.0 && 0.7 & {\bf 12.8}           & {\bf 5.4} &	3.3	& {\bf\phantom{0}7.8}  & --- & 0.4\\
SiS(5-4)                    & 1.0 && 0.7 & \phantom{0}4.0	&	2.7 &	3.9	& \phantom{0}0.8       & 0.3 & 1.1\\
SiS(6-5)                    & 1.0 && 1.0 & \phantom{0}4.7	&	2.2 &	3.8	& {\bf\phantom{0}5.7}  & 1.1 & 0.5\\
HCN                         & 1.0 && 0.8 & \phantom{0}0.9 	&       0.7 &   0.6     &     \phantom{0}0.8   & 0.5 & 0.5\\
H$^{13}$CN(1-0)             & 1.0 && 0.8 & {\bf\phantom{0}6.4}	&	0.9 &	1.8	&     \phantom{0}1.0   & 0.7 & 0.4\\
HNC(1-0)                    & 1.0 && 0.5 & \phantom{0}4.6	&	2.2 &	1.4	&     \phantom{0}3.2   & 0.4 & 0.3\\
SiC$_2$(4,04-3,03)          & 1.0 && --- & ---	                &	2.0 &	1.5	&	---            & --- & ---\\
SiC$_2$(5,05-4,04)          & 1.0 && 1.0 & {\bf\phantom{0}6.9}	&	2.8 &	---	& {\bf 17.7} & --- & ---\\
C$_2$H(1-0)                 & 1.0 && 0.9 & {\bf\phantom{0}5.6}	&	3.0 &	---	&	\phantom{0}2.5 & 2.0 & ---\\
C$_3$H(9/2-7/2)             & 1.0 && 1.1 & \phantom{0}1.7	&	--- &	---	&	---            & --- & ---\\
C$_3$N(11-10)               & 1.0 && 1.3 & \phantom{0}1.5	&	1.6 &	1.2	&	\phantom{0}4.4 & 0.9 & 0.2\\
C$_4$H(10-9)                & 1.0 && 1.2 & {\bf 10.4}           &	2.6 &	---	&	---            & --- & ---\\
C$_3$H$_2$(2,12-1,01)       & 1.0 && --- & {\bf 18.6}           &	--- &	---	&	---            & --- & ---\\
HC$_3$N(10-9)               & 1.0 && 0.6 & \phantom{0}2.4	&	3.0 &	2.2	&	\phantom{0}3.1 & 0.6 & 0.7\\
H$^{(13)}$C$_3$N(10-9)$^a$  & 1.0 && --- & {\bf 16.3}	        &	--- & {\bf 6.4} &	---            & --- & 1.3\\
HC$_3$N(12-11)              & 1.0 && --- & \phantom{0}2.5	&	2.3 &	2.6	&	---            & --- & ---\\
H$^{(13)}$C$_3$N(12-11)$^a$ & 1.0 && 0.4 & {\bf\phantom{0}6.9}	&	--- &	3.3	& {\bf\phantom{0}6.6}  & --- & ---\\
HC$_5$N(34-33)              & 1.0 && --- & ---	                &	--- &	2.2	&	---            & --- & ---\\
\hline
\end{tabular}
 {\footnotesize \\
  \phantom{(a) }Bold face signifies a factor of more than 5.
  \begin{enumerate}
 	\renewcommand{\labelenumi}{(\alph{enumi})}
	\item signifies blend of HCC$^{13}$CN and HC$^{13}$CCN.
  \end{enumerate}
   }
\end{flushleft}
\end{table*}

\begin{table}
   \caption{Fractional abundances for \object{IRC+10216} for species of photospheric origin.}
\label{irc10216_photo}
   \begin{flushleft}
   \begin{tabular}{lcccc}
   \hline\hline
 Species    & Observed$^{\mathrm{a}}$ & & \multicolumn{2}{c}{Model$^{\mathrm{b}}$}  \\ 
\cline{4-5}
                    &                                               & & \multicolumn{1}{c}{TE} & \multicolumn{1}{c}{Shock}
\\

    \hline
CO                   & 1.0(-3)$^{\mathrm{c}}$  && 9.8(-4) & 1.0(-3) \\
HCN                & 1.3(-5)\phantom{$^{\mathrm{c}}$}  && 5.1(-5) & 3.1(-6) \\
CS                   & 9.9(-7)\phantom{$^{\mathrm{c}}$}   && 1.3(-5) & 7.1(-7) \\
SiS                   & 9.5(-7)\phantom{$^{\mathrm{c}}$}  && 1.0(-5) & 2.5(-5)  \\
SiO                   & 1.1(-7)\phantom{$^{\mathrm{c}}$}  && 1.9(-8) & 2.7(-7) \\
  \hline
   \end{tabular}
%
  {\footnotesize
   \begin{enumerate}
        \renewcommand{\labelenumi}{(\alph{enumi})}
        \item From Table~\ref{tababsiz}.
        \item From chemical modelling by \citet{Willacy98}.
        \item Assumed value that determines molecular hydrogen
        densities.
   \end{enumerate}
     }
\end{flushleft}
\end{table}

\section{Discussion}
\label{discuss}
\subsection{Chemistry}
\subsubsection{Presence of shocks in the inner wind?}
When thermal equilibrium (TE) prevails the molecular content of a gas
can be readily calculated from its elemental chemical
composition. This is the case in the stellar photosphere and near the
inner boundary of the envelope of an AGB-star, where the gas density
and temperature are high. The variable nature of AGB-stars induces
pulsation-driven shocks that propagate outwards and suppress TE from
the point of shock formation. Non-equilibrium chemical modelling has
been performed by \citet[][for \object{IRC+10216}]{Willacy98}, and
shows that shocks can strongly alter the chemical abundances in the
inner regions of the CSE from their TE values. In particular, SiO is
strongly enhanced whereas HCN and CS are destroyed. Other species,
e.g., CO and SiS are relatively unaffected.

In Table~\ref{irc10216_photo} the abundances obtained by
\citet{Willacy98} for TE and non-equilibrium chemical modelling (shock
strength of 11.7 km\,s$^{-1}$) are compared to the values obtained in
the analysis. Given the uncertanties, about a factor of five, the
abundances obtained from the observations clearly favour a scenario in
which a shock has passed through the inner ($\lesssim 5$\,R$_{\star}$)
parts of the wind.  The SiO and SiS abundances which are derived are
lower limits since they could be significantly depleted in the outer
envelope due to freeze-out onto dust grains.  The average abundances
derived for HCN, CS, and SiO in the sample are $1\times 10^{-5}$,
$1.5\times 10^{-6}$, $7\times 10^{-7}$, respectively. The SiO
abundance shows the largest spread among the sources reflecting its
sensitivity to the shock strength and possible variation in the
C/O-ratio among the sample sources.  CS might, however, not be
particularly well suited as a probe of shocked
chemistry. \citet{Olofsson_etal1993b} found, when modelling a large
sample of optically bright carbon stars, that in their photospheric
LTE models the CS abundance varied considerably with adopted stellar
temperature.

\subsubsection{Photochemistry in the outer envelope}

Many photochemical models have been developed for the outer envelope
of \object{IRC+10216} \citep[e.g.][]{MillarHerbst1994, Doty98,
Millar_etal2000}. The physical conditions in the outer parts of the
wind ($\gtrsim 100$\,R$_{\star}$) allow for the penetration of ambient
ultraviolet radiation that induces a photochemistry.  In
Table~\ref{irc10216} the derived column densities for
\object{IRC+10216} for a number of species produced in the envelope
are compared to those from the photochemical model of
\citet{Millar_etal2000}. In general the abundances agree well, given
the uncertainties.

It is remarkable that the abundances generally show relatively little
variation within the sample.  Most apparent is the over-abundance of
Si-bearing molecules in \object{CIT 6} (Sect.~\ref{cit6}). Some of the
abundances of \object{IRAS 15194--5115} also stand out and will be
separately discussed in Sect.~\ref{iras15194}.  There are no apparent
trends with the stellar or circumstellar parameters. This would
suggest that the physical structure in these sources indeed is much
the same and that the initial atomic abundances are similar. When
comparing the present sample of carbon stars to the sample of
\citet{Olofsson_etal1993b}, which on the whole tend to have a low
mass-loss rate ($\sim$10$^{-7}$\,M$_\odot$\,yr$^{-1}$), the agreement
in derived abundances for the star in common, \object{CIT 6}, is very
good. There is less than a factor 3 difference in the calculated
abundances of HCN, CN and CS. However, in general, the sample of low
mass-loss rate stars has calculated abundances which are
systematically an order of magnitude greater than those calculated
here.  It must be noted that \citet{Olofsson_etal1993b} use an
excitation temperature of 20\,K, twice that used in this
analysis. Moreover, \citet{SchoierOlofsson2001} show that the
mass-loss rates in the low mass-loss rate objects in
\citet{Olofsson_etal1993b} are underestimated by about a factor of 5
on average. This would explain the apparent discrepancy in calculated
abundances between the high mass-loss rate objects here and the lower
mass-loss rate objects in \citet{Olofsson_etal1993b}. Hence there
seems not to be a marked difference in the molecular composition of
high and low mass-loss carbon stars.

The large spread in the abundance of isotopomers containing $^{13}$C
follows from the varying $^{12}$C/$^{13}$C-ratio among the sample
sources. The $^{12}$C/$^{13}$C-ratio is related to the nucleosynthesis
rather than the chemistry and reflects the evolutionary status of
these stars.  However, chemical fractionation may affect this ratio in
certain molecules, in particular CO.

CN/HCN and HNC/HCN ratios can also be used to estimate the
evolutionary status of carbon stars. CN is produced via the
photodissociation of HCN by ultraviolet radiation, and as stellar
radiation increases with evolution from AGB star to PPN to PN, so the
CN/HCN ratio will increase
\citep[e.g.,][]{Bachiller_etal1997b,Cox_etal1992}. HNC, formed from
the dissociative recombination of HCNH$^+$, behaves in a similar
way. In this sample there is a rather large spread in the CN/HCN
ratio, from 0.07--0.68, and a value of 1.82 for \object{CIT 6} (see
Sect. \ref{cit6}). This large spread is in contrast to
\citet{Olofsson_etal1993b}, for example, who found a very narrow range
(CN/HCN $\sim$0.65--0.70) in low mass-loss stars. However, there is
excellent agreement with the results of
\citet{Lindqvist_etal2000}. For the carbon stars \object{IRC+10216},
\object{CIT 6} and \object{IRC+40540} they derive ratios of 0.16, 1.5
and 0.17, respectively, in comparison with the 0.22, 1.82 and 0.15
derived here. A CN/HCN ratio of $\sim$0.5 is typical in C-rich AGB
stars \citep{Bachiller_etal1997a}, increasing to $\sim$5 in PPNe. In
fact, a ratio of 0.6--0.7 is predicted by the photodissociation model,
with only a weak dependence on mass-loss rate \citep[see Fig. 8
of][]{Lindqvist_etal2000}. The HNC/HCN ratio seems to be split into
two ranges in the present sample of carbon stars. \object{IRC+10216},
\object{AFGL 3068} and \object{IRC+40540} have HNC/HCN ratios of
$\leq$0.005, whilst the remaining stars have ratios of 0.01--0.03. The
value derived for \object{IRC+10216} is in agreement with that quoted
in \citet{Cox_etal1992}. This seems to indicate that the sample stars
are not well evolved, since a HNC/HCN ratio of $\sim$1 is expected in
PPNe \citep[e.g.,][]{Cox_etal1992}. Having said this, the HNC/HCN
ratio does rapidly become of the order 1, as can be seen in models of
PPN chemistry \citep{Woods_etal2002}.

Generally, it seems that to use the term \textquotedblleft carbon
chemistry\textquotedblright\ to refer to a paradigm of chemistry in
C-rich evolved stars is reasonable. Of the sample stars here, given
the variety of molecular species, there is very little difference in
molecular abundances, save for two slightly curious sources, as
detailed in the following subsections.

\subsection{IRAS 15194--5115}
\label{iras15194}
Of the derived abundances those obtained for \object{IRAS 15194--5115}
stand out the most.  In particular, the SiO and C$_4$H abundances, in
addition to the isotopomers containing $^{13}$C, appear significantly
enhanced towards this source. C$_3$H$_2$ also appears to be greatly
enhanced in this source, but, however, this molecule is only observed
in one other star, \object{IRC+10216}.  The $^{12}$CO/$^{13}$CO-ratio
of 6 derived for \object{IRAS 15194--5115} is significantly lower than
that of the others and that which is commonly derived for carbon
stars. This value is certain, with the modelling of the CO
emission being supported by intensity ratios for another four species,
which agree to $\pm$30\% (Table~\ref{12c/13c}). The evolutionary
status of this star is undetermined.  \citet{Ryde_etal1999} speculated
that \object{IRAS 15194--5115} might be a massive (5--8\,M$_{\sun}$)
star in the last stages of evolution where its low
$^{12}$C/$^{13}$C-ratio is the result of hot bottom burning (HBB). The
increase of $^{14}$N from the CNO cycle is a signature of HBB
\citep{Marigo2001, Ventura_etal2002}, but given the uncertainty in the
data presented here, this suggestion cannot be confirmed.

\subsection{CIT 6}
\label{cit6}
\object{CIT 6} is another object outstanding in the sample. It has a
CN/HCN ratio of $\sim$1.8, which suggests an advanced evolutionary
status, but a low HNC/HCN ratio ($\sim$0.02), which suggests the
contrary. Certainly the idea that \object{CIT 6} is well on its way to
becoming a PPN has been put forward before
\citep[e.g.,][]{Trammell_etal1994,Monnier_etal2000,Zacs_etal2001}.

The modelled $^{12}$CO/$^{13}$CO ratio in this source agrees well with
that carried out previously
\citep{Groen_etal1996,SchoierOlofsson2000}. This ratio, however, does
not agree with $^{12}$C/$^{13}$C ratios derived from observations of
other molecules and their $^{13}$C isotopes, both in this paper
(Table~\ref{12c/13c}) and elsewhere \citep{Kahane92,Groen_etal1996}.

A further point worth note is the comparative over-abundance of
Si-bearing species which possibly indicates a less efficient
freeze-out onto dust grains in this particular source.

\begin{table}
   \caption{Radial column densities (cm$^{-2}$) for species of circumstellar origin towards \object{IRC+10216}.}
\label{irc10216}
   \begin{flushleft}
   \begin{tabular}{lcccc}
   \hline\hline
 Species    & Observed$^{\mathrm{a}}$ & & MHB$^{\mathrm{b}}$ & Obs./MHB  \\ 
    \hline
CN                 & 8.3(14) && 1.0(15) & 0.8\\
HNC                & 2.0(13) && 8.4(13)  & 0.2\\
C$_2$H             & 8.9(14) && 5.7(15) & 0.2\\
C$_3$H             & 2.1(13) && 1.4(14) & 0.2\\
C$_3$N             & 2.0(14) && 3.2(14) & 0.6\\
C$_4$H             & 9.3(14) && 1.0(15) & 0.9\\
C$_3$H$_2$         & 1.2(13) && 2.1(13)  & 0.6\\
HC$_3$N            & 9.1(14) && 1.8(15) & 0.5\\
CH$_3$CN           & 9.9(12) && 3.4(12) & 2.9\\
HC$_5$N            & 5.8(15) && 7.1(14) & 8.2\\
  \hline
   \end{tabular}
%
  {\footnotesize
   \begin{enumerate}
        \renewcommand{\labelenumi}{(\alph{enumi})}
        \item Calculated from Tables~\ref{tabmlr} \& \ref{tababsiz}.
        \item From chemical modelling by \citet{Millar_etal2000}.
   \end{enumerate}
     }
\end{flushleft}
\end{table}

\section{Conclusions}
The seven high mass-loss rate carbon stars presented here exhibit rich
spectra at millimetre wavelengths with many molecular species readily
detected. A total of 47 emission lines from 24 molecular species were
detected for the sample stars.  The mass-loss rate and physical
structure of the circumstellar envelope, such as the density and
temperature profiles, was carefully estimated based upon a detailed
radiative transfer analysis of CO. The determination of the mass-loss
rate enables abundances for the remaining molecular species to be
calculated. The derived abundances typically agree within a factor of
five indicating that circumstellar envelopes around carbon stars have
similar molecular compositions.

The most striking difference between the abundances are reflecting the
spread in the $^{12}$C/$^{13}$C-ratio of about an order of magnitude
between the sample stars.  Also, the high abundance of SiO in the
envelopes indicates that a shock has passed through the gas in the
inner parts of the envelope. This is further corroborated by the
relatively low amounts of CS and possibly HCN.

The abundances of species that are produced in the outer parts of the
wind can be reasonably well explained by current photochemical models.

\begin{acknowledgements}
We are grateful to F.~Kerschbaum for help with estimating some of the
input parameters to the CO modelling.  FLS is supported by the
Netherlands Organization for Scientific Research (NWO) grant
614.041.004. The Swedish--ESO Submillimetre Telescope, SEST, is
operated jointly by ESO and the Swedish National Facility for
Radioastronomy, Onsala Space Observatory at Chalmers University of
Technology. The OSO 20\,m telescope is operated by the Swedish
National Facility for Radioastronomy. This article made use of data
obtained through the JCMT archive as Guest User at the Canadian
Astronomy Data Center, which is operated by the Dominion Astrophysical
Observatory for the National Research Council of Canada's Herzberg
Institute of Astrophysics.
      
\end{acknowledgements}


\bibliographystyle{aa}

\appendix

\section{Spectra}

   \begin{figure*} \centering{
   \includegraphics[width=16.75cm]{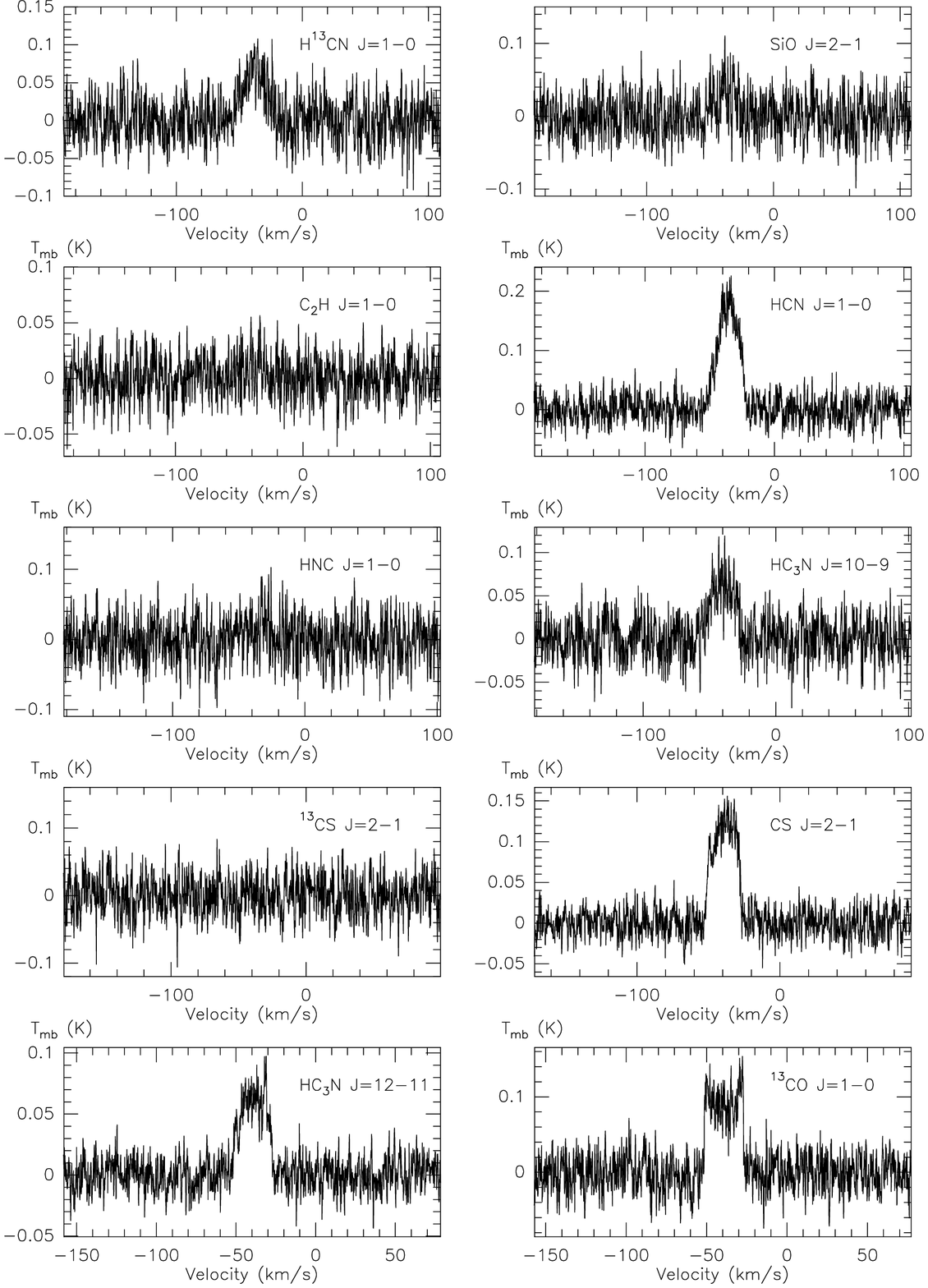}
   \caption{High-resolution spectra of \object{IRAS07454--7112}, obtained
   with the SEST.}  \label{07454hrs1}} \end{figure*}

   \begin{figure*}
   \centering{   
   \includegraphics[width=16.75cm]{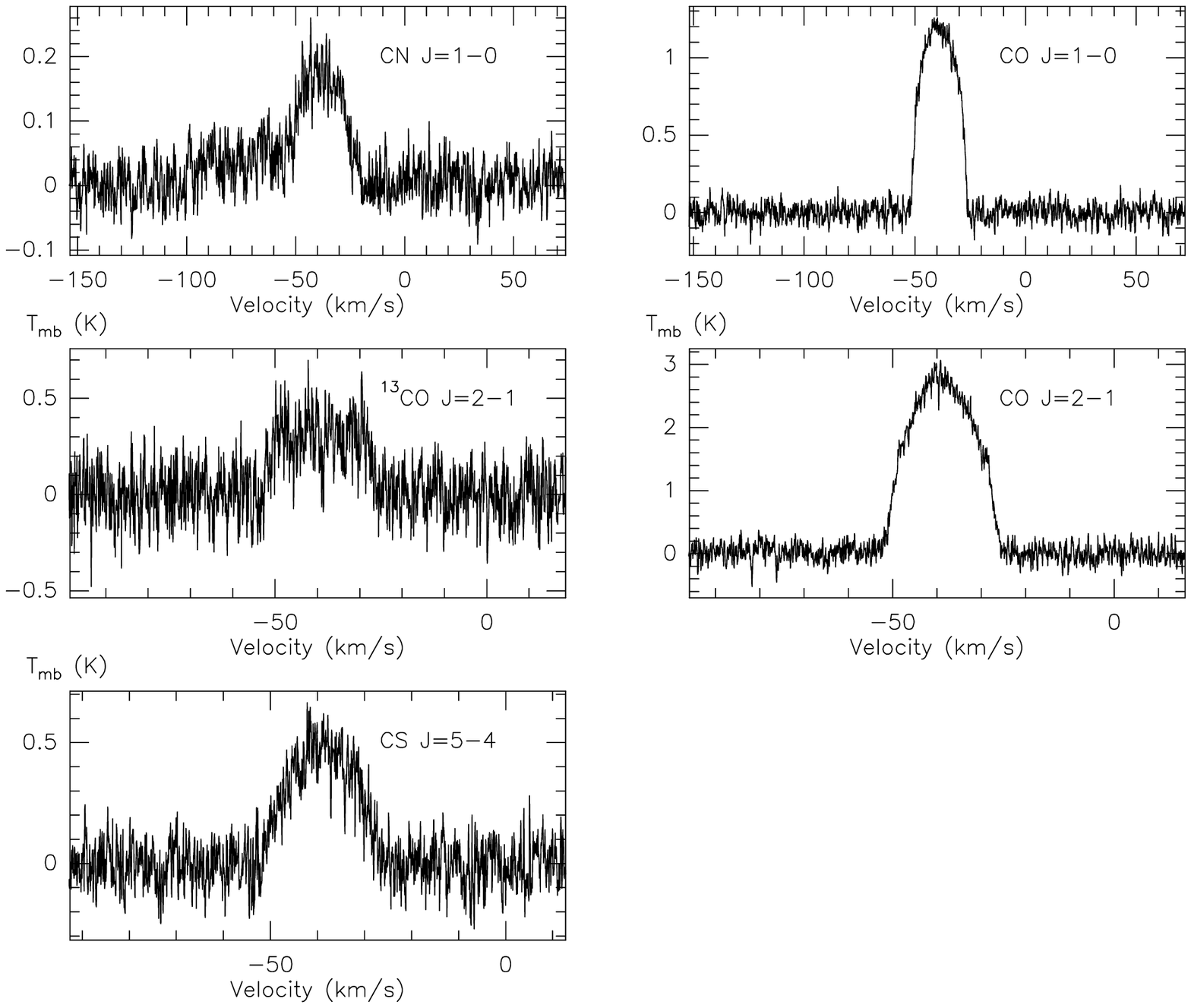}
   \caption{High-resolution spectra of \object{IRAS07454--7112}, obtained
   with the SEST.}
   \label{07454hrs2}}
   \end{figure*}

   \begin{figure*}
   \centering{   
   \includegraphics[width=16.75cm]{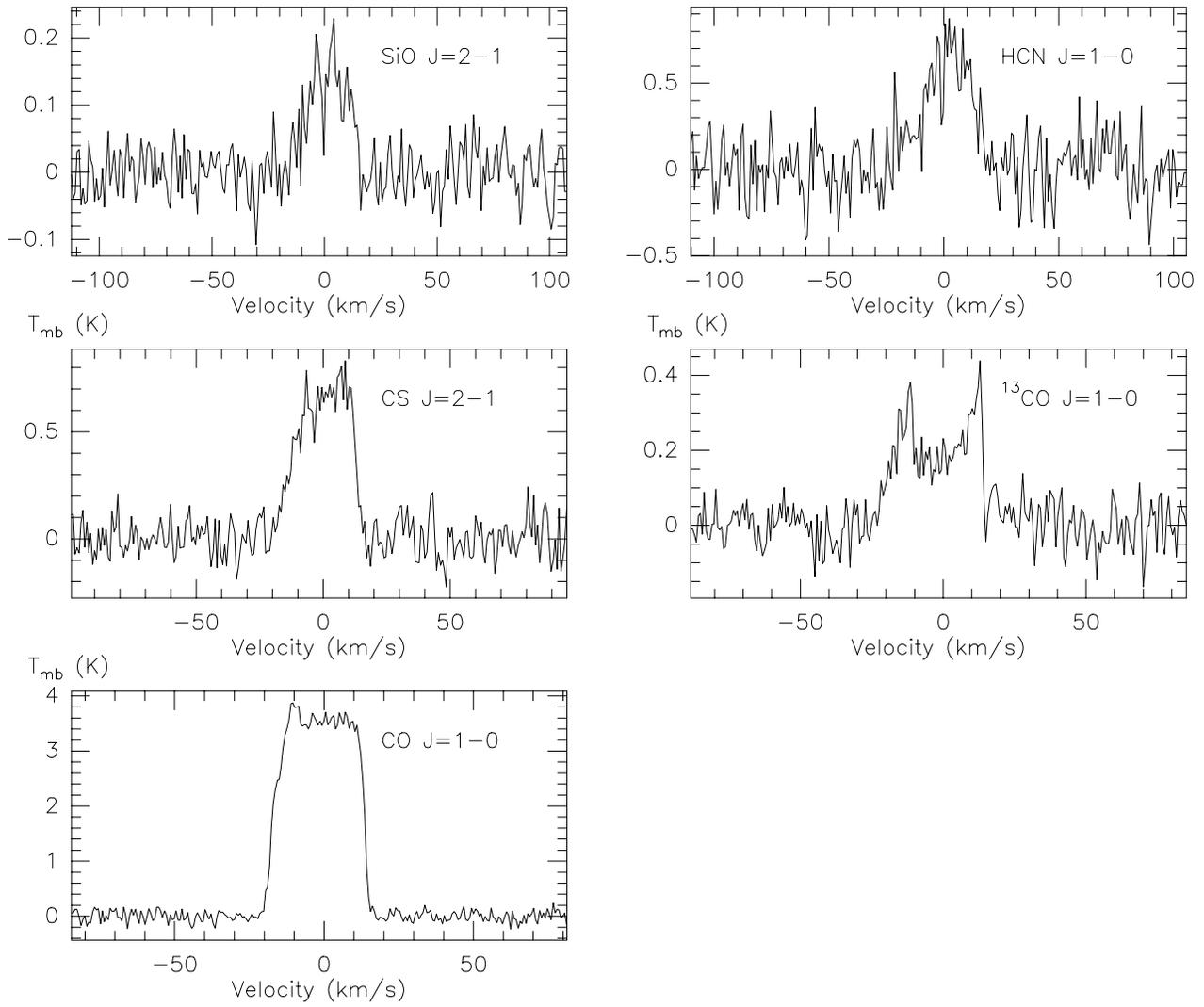}
   \caption{High-resolution spectra of \object{CIT\,6}, obtained
   at OSO.}
   \label{cit6hrs1}}
   \end{figure*}

   \begin{figure*}
   \centering{   
   \includegraphics[width=16.75cm]{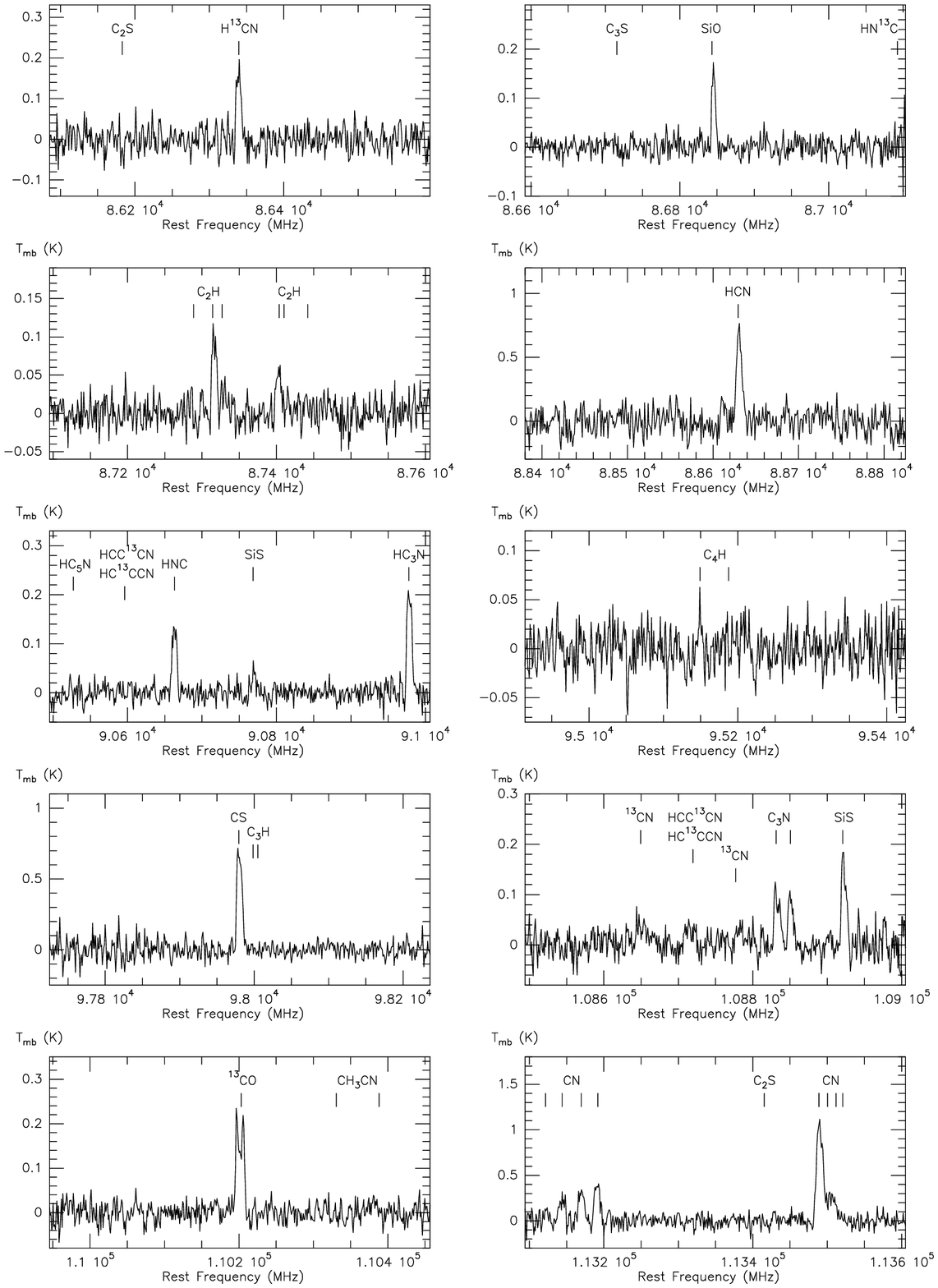}
   \caption{Low-resolution spectra of \object{CIT\,6}, obtained
   at OSO.}
   \label{cit6lrs1}}
   \end{figure*}

   \begin{figure*}
   \centering{   
   \includegraphics[width=16.75cm]{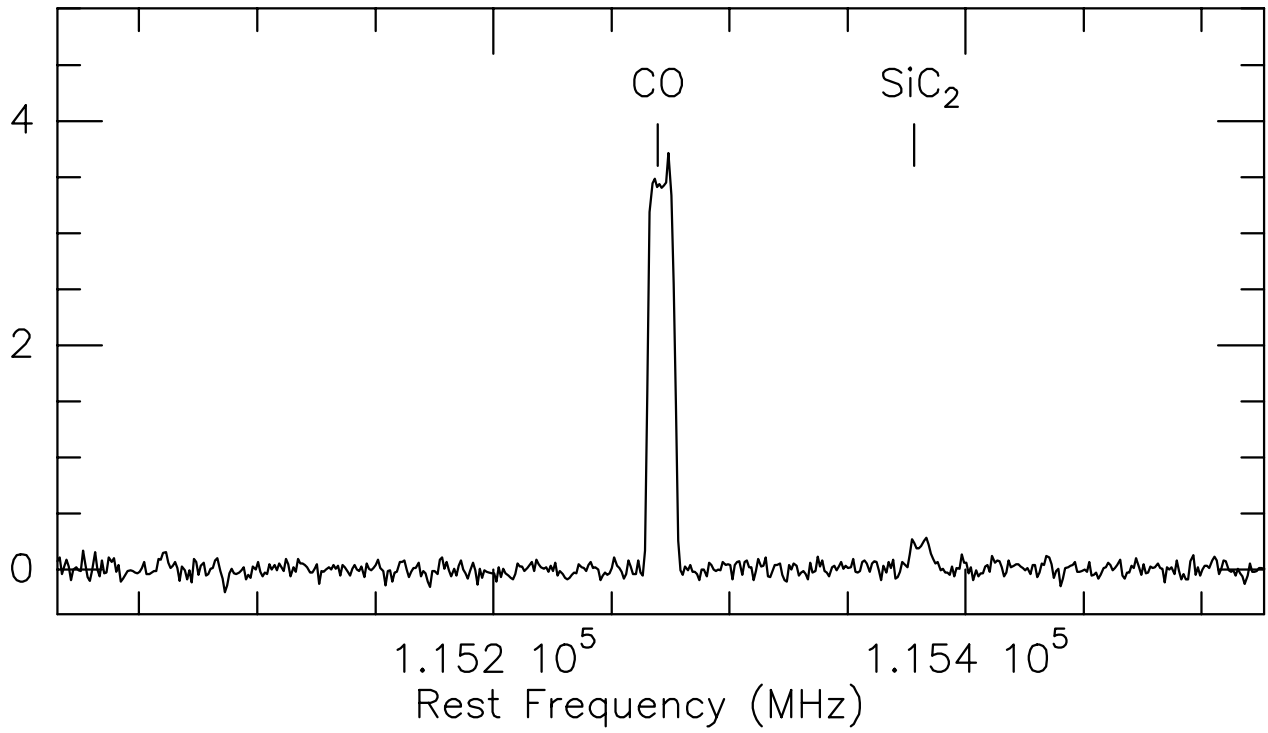}
   \caption{Low-resolution spectrum of \object{CIT\,6}, obtained
   at OSO.}
   \label{cit6lrs2}}
   \end{figure*}

   \begin{figure*}
   \centering{   
   \includegraphics[width=16.75cm]{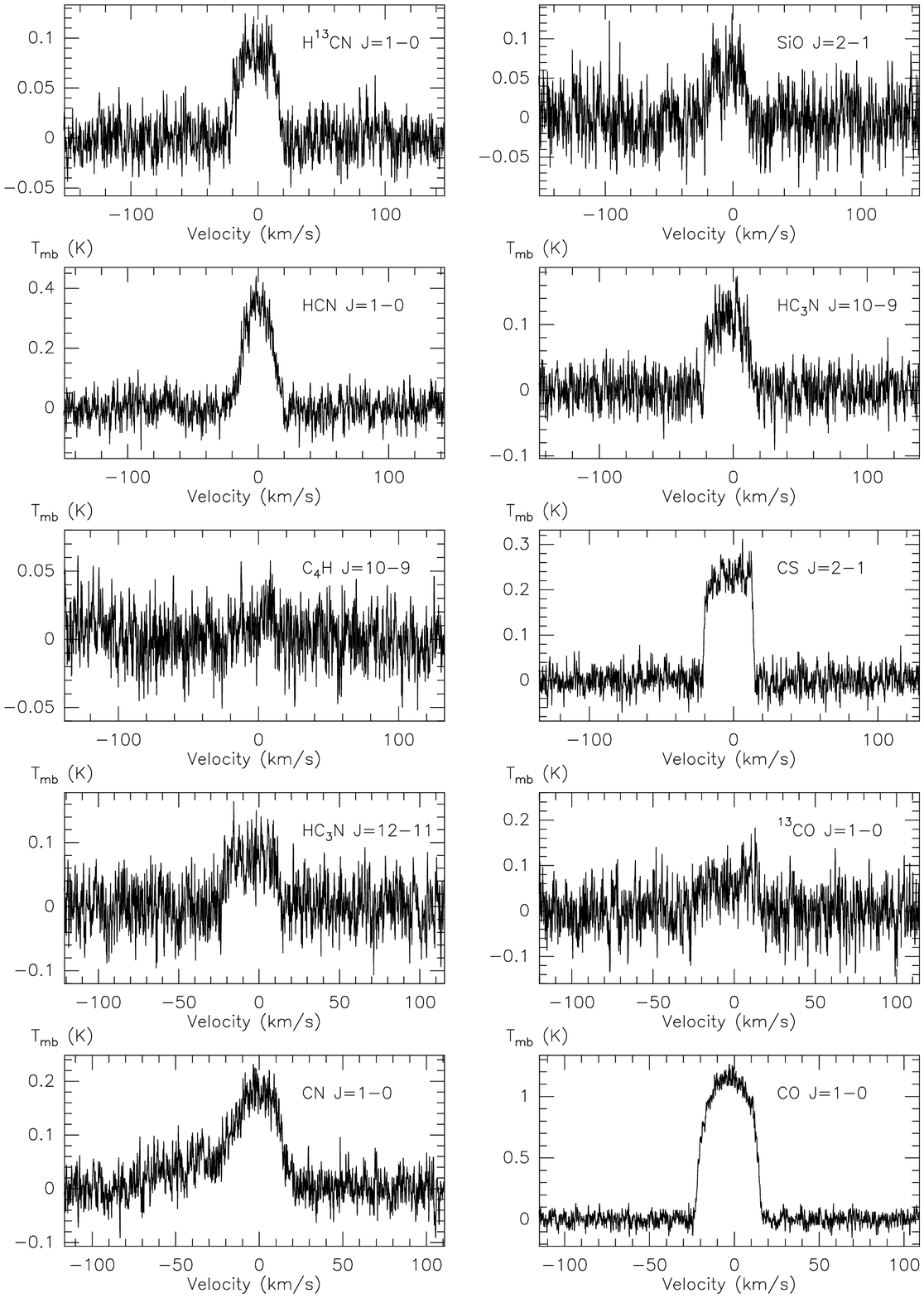}
   \caption{High-resolution spectra of \object{IRAS15082--4808}, obtained
   with the SEST.}
   \label{15082hrs1}}
   \end{figure*}

   \begin{figure*}
   \centering{   
   \includegraphics[width=16.75cm]{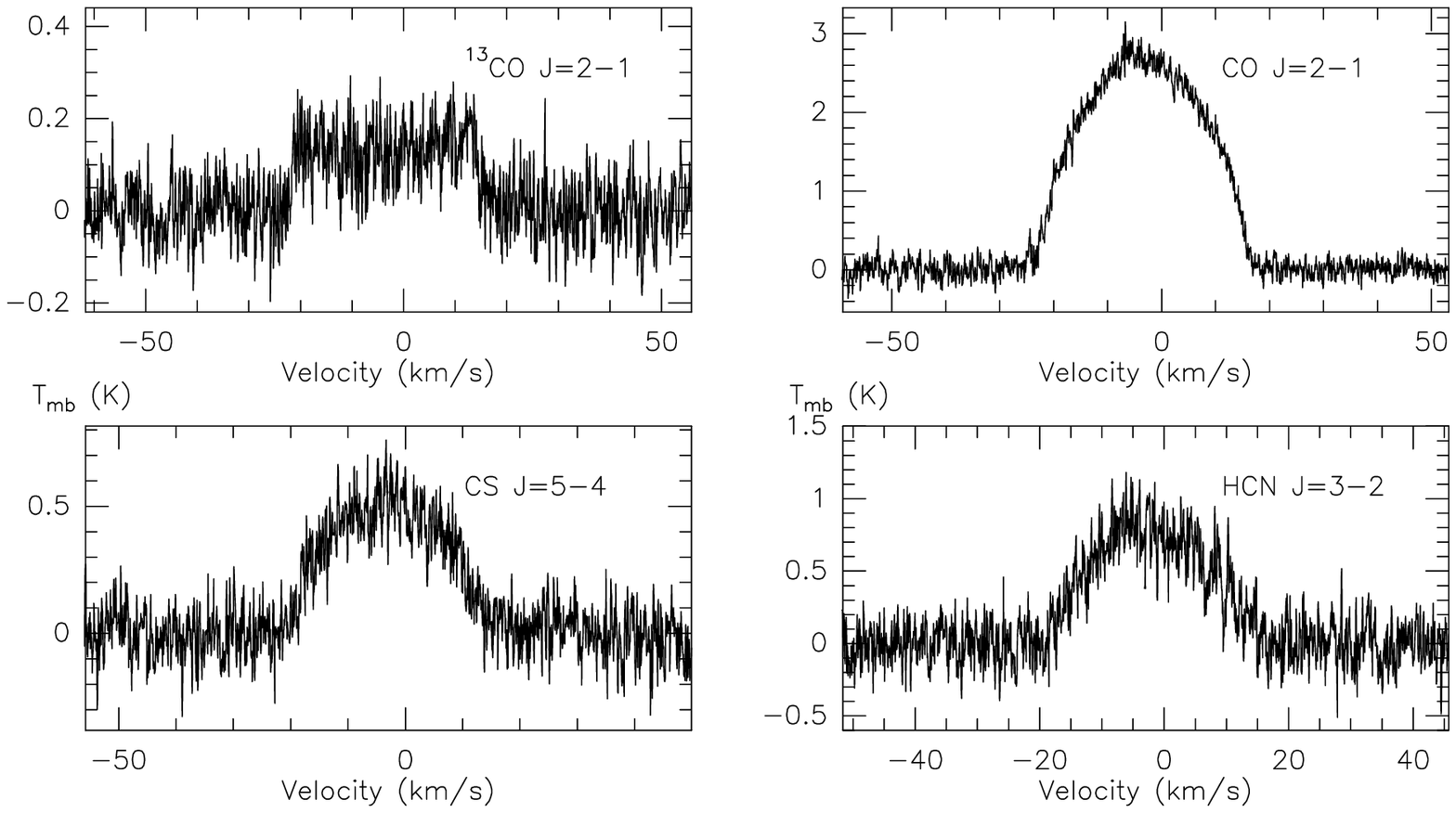}
   \caption{High-resolution spectra of \object{IRAS15082--4808}, obtained
   with the SEST.}
   \label{15082hrs2}}
   \end{figure*}

   \begin{figure*}
   \centering{   
   \includegraphics[width=16.75cm]{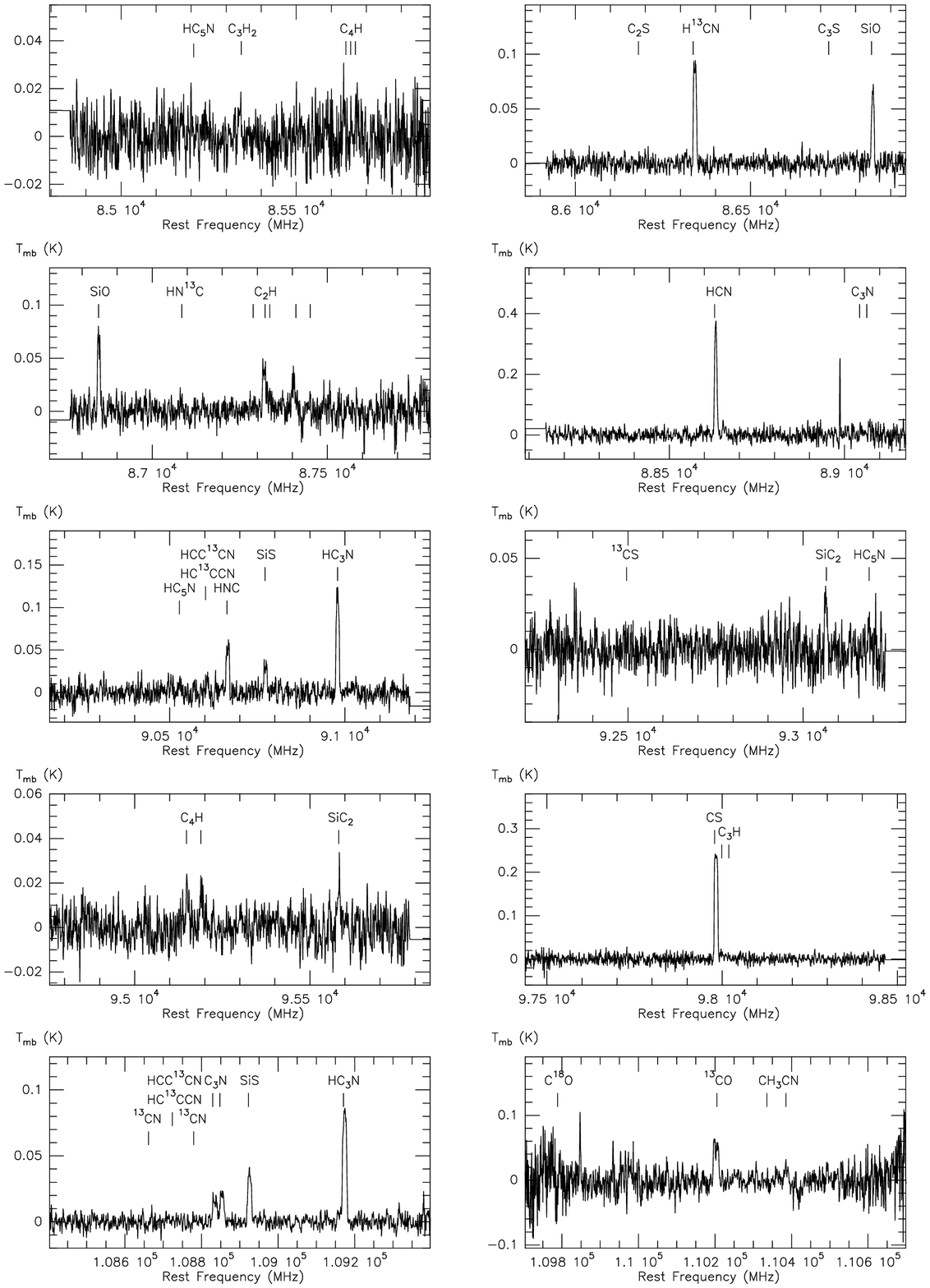}
   \caption{Low-resolution spectra of \object{IRAS15082--4808}, obtained
   with the SEST.}
   \label{15082lrs1}}
   \end{figure*}

   \begin{figure*}
   \centering{   
   \includegraphics[width=16.75cm]{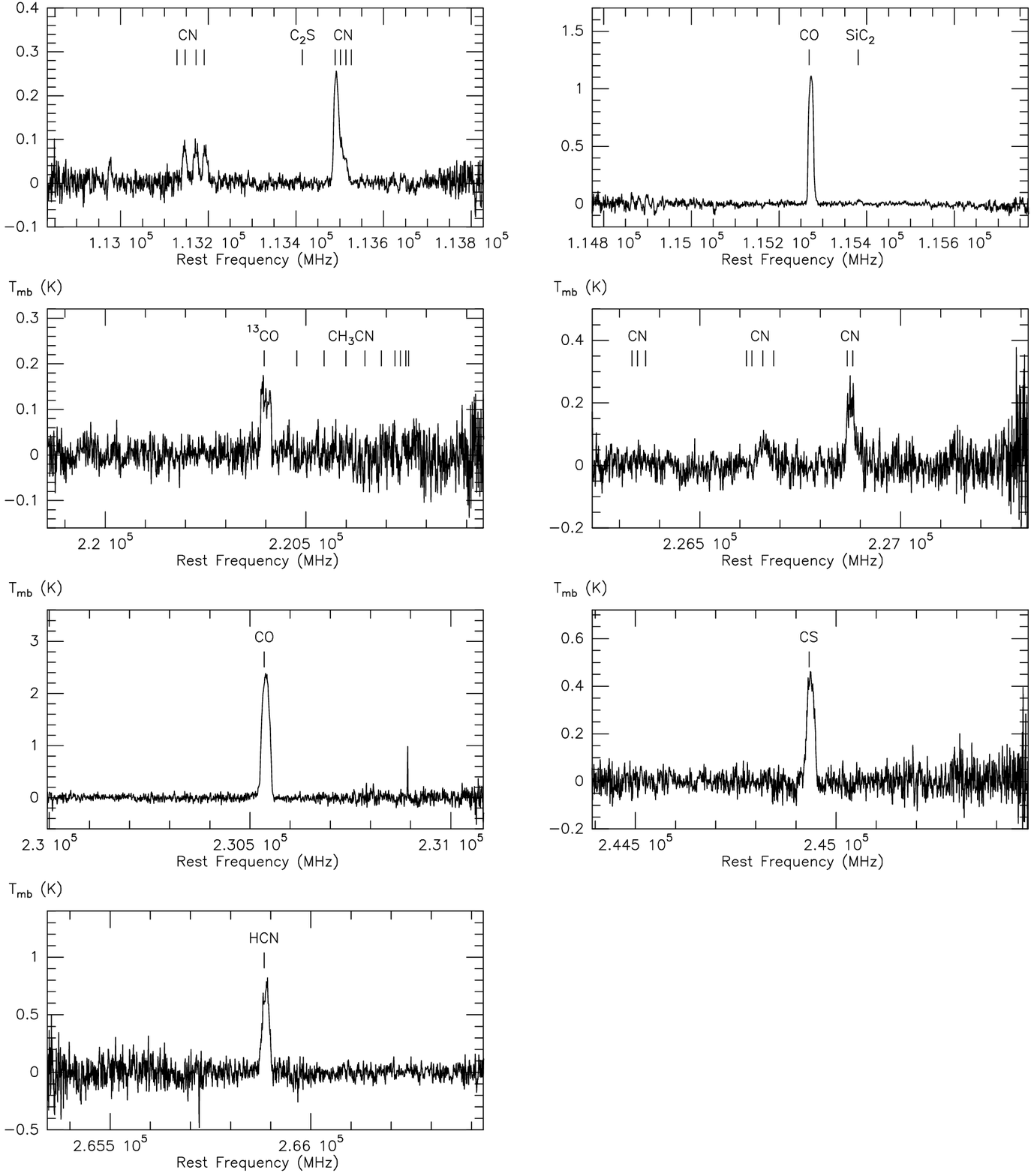}
   \caption{Low-resolution spectra of \object{IRAS15082--4808}, obtained
   with the SEST.}
   \label{15082lrs2}}
   \end{figure*}

   \begin{figure*}
   \centering{   
   \includegraphics[width=16.75cm]{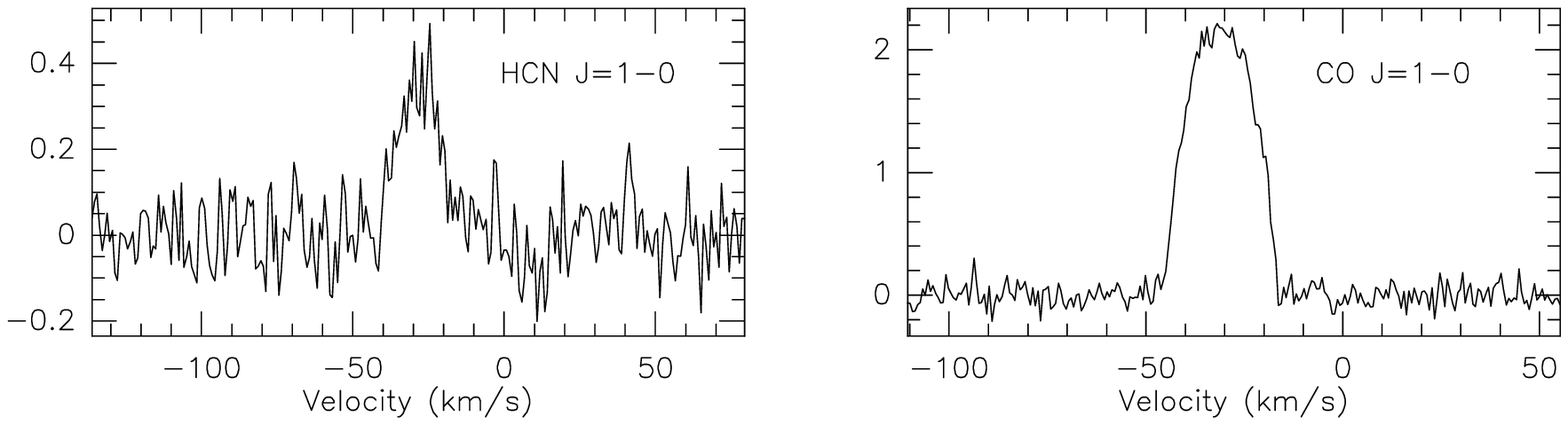}
   \caption{High-resolution spectra of \object{AFGL3068}, obtained
   at OSO.}
   \label{3068hrs1}}
   \end{figure*}

   \begin{figure*}
   \centering{   
   \includegraphics[width=16.75cm]{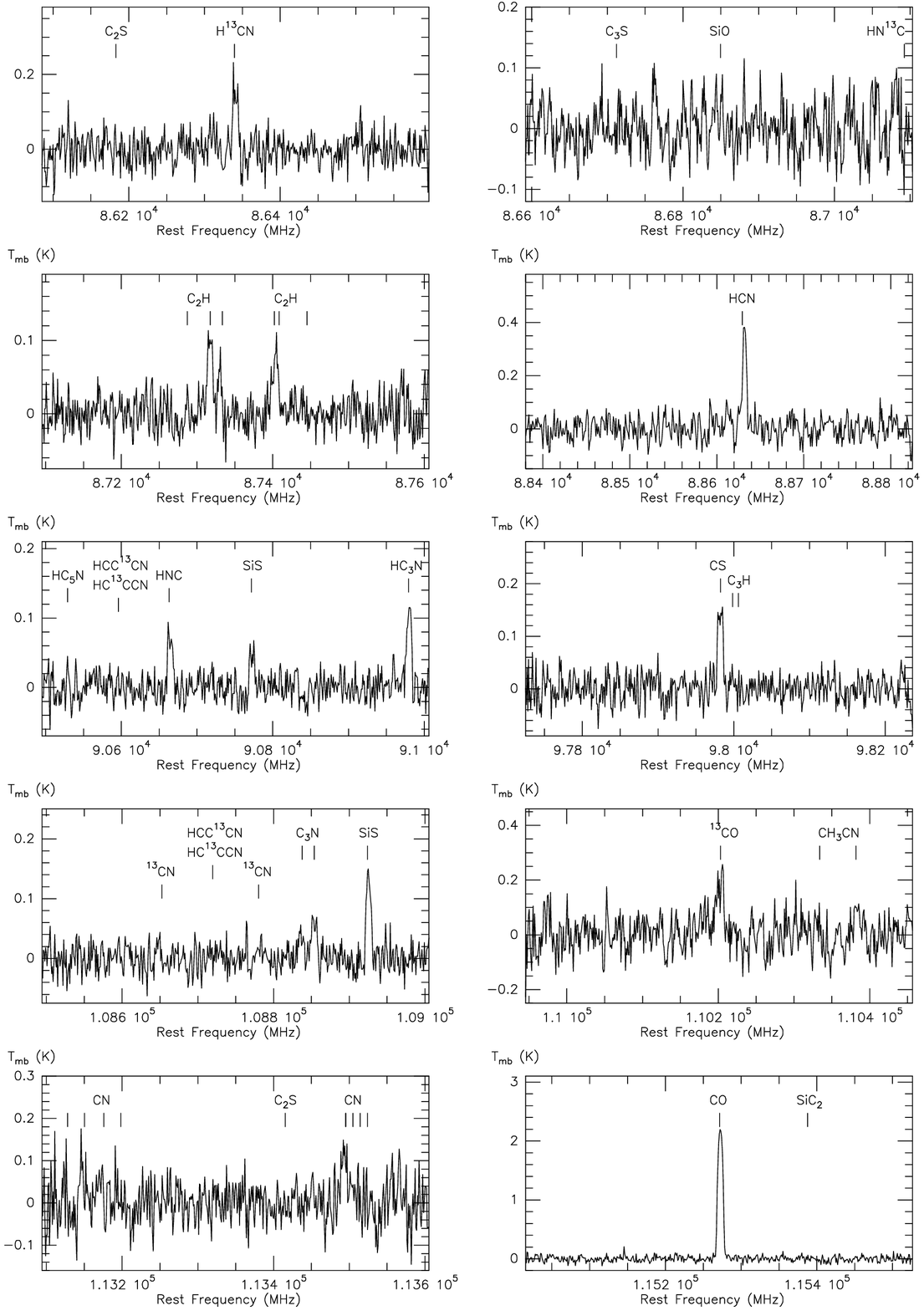}
   \caption{Low-resolution spectra of \object{AFGL3068}, obtained
   at OSO.}
   \label{3068lrs1}}
   \end{figure*}

   \begin{figure*}
   \centering{   
   \includegraphics[width=16.75cm]{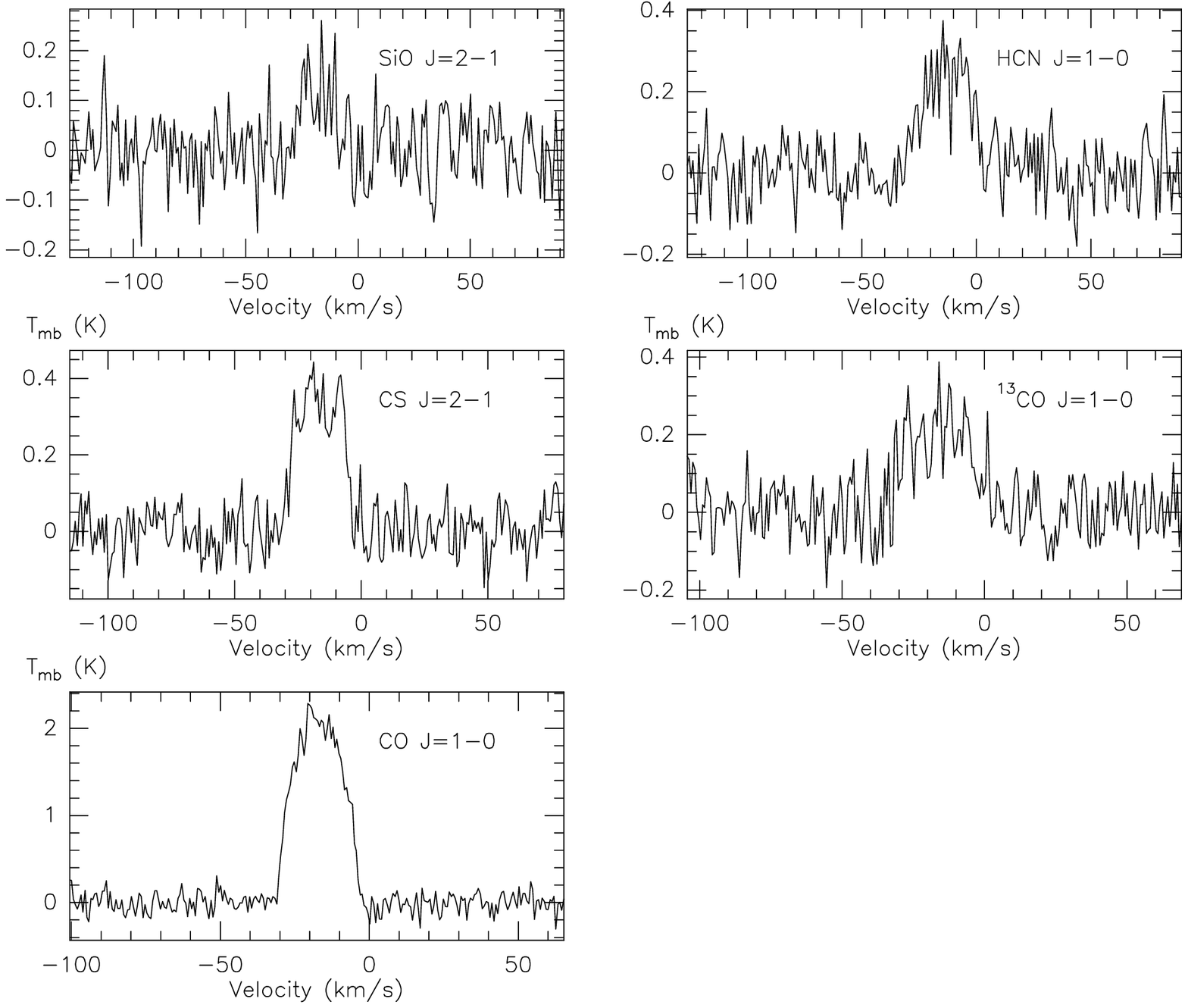}
   \caption{High-resolution spectra of \object{IRC+40540}, obtained
   at OSO.}
   \label{40540hrs1}}
   \end{figure*}

   \begin{figure*}
   \centering{   
   \includegraphics[width=16.75cm]{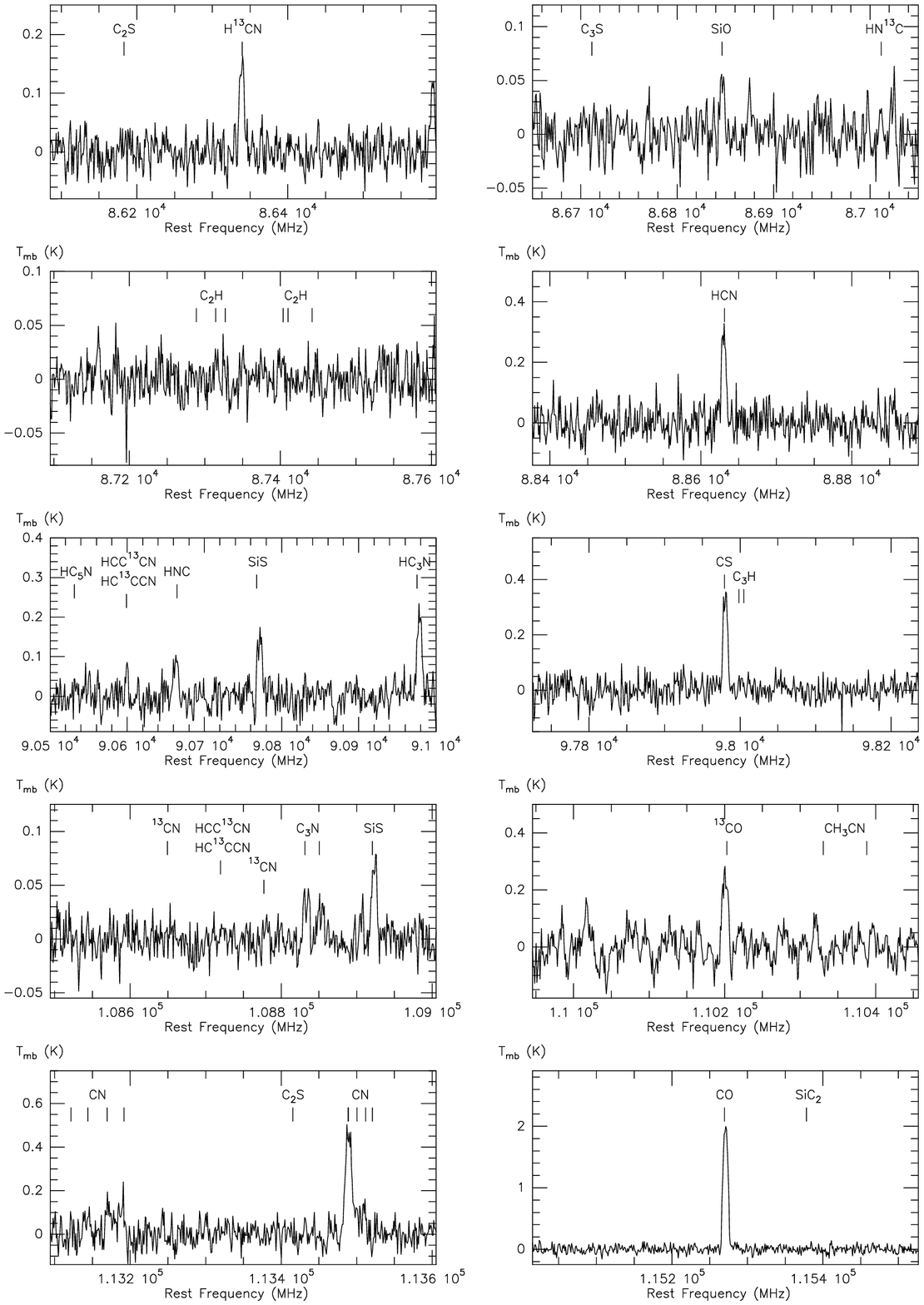}
   \caption{Low-resolution spectra of \object{IRC+40540}, obtained
   at OSO.}
   \label{40540lrs1}}
   \end{figure*}

\end{document}